\documentclass[sigconf]{acmart}
\AtBeginDocument{%
  \providecommand\BibTeX{{%
    \normalfont B\kern-0.5em{\scshape i\kern-0.25em b}\kern-0.8em\TeX}}}

\setcopyright{none}

\acmConference[WSDM'22]{WSDM'22: ACM International Conference on Web Search and Data Mining}{February 21--25, 2022}{Phoenix, Arizona}
\usepackage{subcaption}
\usepackage{caption}
\usepackage{graphicx}
\usepackage{multirow}
\usepackage{makecell}
\graphicspath{{figures/}}

\begin{document}

\title[Neural Mode Decomposition]{Neural Mode Decomposition based on Fourier neural network and frequency clustering}

\author{Yiting Hu}
\orcid{1234-5678-9012}
\affiliation{%
  \institution{Beihang University}
  \department{School of Electronic and Information Engineering}
  \streetaddress{Haidian Qu}
  \city{Beijing}
  \country{China}
}
\email{huyiting@buaa.edu.cn}

\author{Zhuangzhi Wu}
\affiliation{%
  \institution{Beihang University}
  \department{School of Computer Science and Engineering}
  \streetaddress{Haidian Qu}
  \city{Beijing}
  \country{China}
}
\email{zzwu@buaa.edu.cn}
\renewcommand{\shortauthors}{Hu and Wu.}

\begin{abstract}
Since Huang proposed the Empirical Mode Decomposition (EMD) in 1998, mode decomposition has been widely studied, but EMD and relative developed algorithms are still generally lack of adaptability and mathematical theory. This paper propose a new mode decomposition algorithm called Neural Mode Decomposition (NMD) based on Fourier neural network (FNN) and frequency clustering. Firstly, a FNN is constructed to decompose and learn the information of each amplitude modulation frequency component and non-periodic component in the raw data. Secondly, the frequency components obtained by the FNN are clustered into multiple Intrinsic Mode Functions (IMF) with separated spectrum based on the energy of each frequency component learned by FNN. Practical decomposition results on a series of artificial and real data show that NMD algorithm can effectively implement mode decomposition, better reflect the characteristics of raw data than EMD, and has higher adaptability than Variational Mode Decomposition (VMD).
\end{abstract}



\maketitle

\section{Introduction}
When studying time varying signals, we often consider the signal decomposition technique with time-frequency characteristics, in which the original complex irregular signals are decomposed into several simple and regular components for exploring its deep pattern. In general, time-frequency decomposition methods can be divided into two categories: decomposition with pre-assigned family of templates and data driven decomposition. In the first category, examples are windowed Fourier transform \cite{Gabor1946TheoryOC} and wavelet transform \cite{Daubechies1988}, where the templates are respectively generated from the trigonometric functions and the basic wavelet. In the second category, mode decomposition is a large family of highly adaptive methods originated from EMD \cite{huang1998empirical}, where the raw data are decomposed into the sum of Intrinsic Mode Function (IMF) and a non-periodic residual. EMD is not based on the templates but based on the raw data itself, has a wider and more flexible application than decomposition methods with pre-assigned templates.

In mode decomposition, the IMF needs to satisfy the single mode condition at one time point, so that the instantaneous frequency of a IMF is of physical significance for time-frequency analysis. In \cite{huang1998empirical}, IMF is defined by two conditions. The first is that the difference between the number of local extrema and zero-crossings is no more than one, and the second is that the average of the envelopes of local maximum point and local minimum point must be zero at any time point. EMD can be widely used in different kinds of data, but its decomposition results are highly affected by extreme point detection, the method of finding envelopes, and the stopping criteria. Moreover, it is sensitive to noise and lack of mathematical theory. Hence, in order to give IMF a clearer interpretation, \cite{daubechies2011synchrosqueezed} improve the definition of IMF to a special amplitude-modulated-frequency-modulated (AM-FM) signal:
\begin{equation}\label{Eq1}
u_{k}(t)=a_{k}(t)\cos{\phi_{k}(t)}, a_{k}(t), \phi_{k}^{\prime}(t)>0, \forall(t)
\end{equation}
where $u_{k}(t)$ is the $k$th decomposed IMF, $\phi_{k}(t)$ is a nondecreasing function, amplitude function $A_{k}(t)$ is nonnegative, and $a_{k}(t)$ and $\phi_{k}^{\prime}(t)$ vary much slower than $\phi_{k}(t)$. That is, in a sufficiently long interval, the mode can be considered to be a pure harmonic signal with amplitude $a_{k}(t)$ and instantaneous frequency $\phi_{k}^{\prime}(t)$ \cite{daubechies2011synchrosqueezed}. The definition of IMF based on AM-FM signal satisfies the definition of IMF in \cite{huang1998empirical}, but the converse is not necessarily true.

Based on the developed definition of IMF, several mode decomposition method differing from the recursive EMD have been proposed. In \cite{gilles2013empirical}, the author introduce the Empirical Wavelet Transform (EWT) combining the characteristics of compact support Fourier spectrum of AM-FM components with wavelet analysis, which can build adaptive wavelets, segment the Fourier spectrum and obtain a set of orthonormal components. The decomposition results of EWT are more consistent with the origin data than EMD, and have better interpretation. Dragomiretskiy \cite{dragomiretskiy2013variational} introduce Variational Mode Decomposition (VMD) based on the AM-FM-signal-like IMF. VMD take the sum of the estimated bandwidth of each IMF as the objective function. The decomposition which minimize the objective function is the final mode decomposition results, in which each IMF is compact around a central frequency. VMD can overcome the disadvantage of mode mixing in EMD, with better orthogonality and noise robustness, and has been widely used in data decomposition. However, VMD needs to predetermine the number of modes, which greatly impact the results. \cite{lian2018adaptive} develop the Adaptive Variational Mode Decomposition (AVMD) to automatically determine the mode number, solving the drawbacks that the number of modes needs to be manually selected in VMD. In \cite{ur2019multivariate}, the application scenario of VMD is extended from univariate signal to multivariate data. Based on the principal that a wide-band nonlinear chirp signal (NCS) can be transformed to a narrow-band signal utilizing demodulation techniques, \cite{chen2017nonlinear} focus on wide-band NCSs and propose Variational Nonlinear Chirp Mode Decomposition (VNCMD) combining VMD with the optimal demodulation problem. Also, for chirp signals, \cite{chen2017intrinsic} propose Intrinsic Chirp Component Decomposition (ICCD) modeling the instantaneous frequencies and instantaneous amplitudes of the intrinsic chirp components as Fourier series and obtain the decomposition when reconstructing the original data. \cite{chen2020multivariate} and \cite{chen2021multivariate} extend the application of VNCMD and ICCD from univariate signal to multivariate data respectively. Most of the researches mentioned above include complex objective functions and algorithms, with certain limits in the practical application. Compared with the classical EMD, these methods have better mathematical theory, in which the characteristics of decomposition results are more regular and significant, but reduce the adaptability at the same time. Hence, we still need a mode decomposition method with both mathematical theory and adaptability, which can effectively extract data features.

Artificial neural network (ANN) is one of the most popular technologies in the 21st century. It constructs the combination of multiple neurons with active functions, affording to fit a variety of multi-dimension complex functions and having high adaptability. We believe that the application of ANN in signal decomposition can break through the limitations of current mode decomposition methods and obtain more satisfactory decomposition results. 
Fourier neural network (FNN) is a kind of ANN taking trigonometric as active functions, based on the theory of Discrete Fourier Transform (DFT), which can obtain DFT and Inverse Discrete Fourier Transform (IDFT) \cite{gashler2016modeling}. Some FNN can modify the given harmonic frequency and phase in DFT \cite{godfrey2017neural}, so that the results are more consistent to the original data. Since the neuron with trigonometric active function and AM-FM-like IMF have similarity on structure, 
this paper propose a mode decomposition model Neural Mode Decomposition (NMD) based on FNN.

The remainder of this paper is organized as follows. Section \ref{sec2} gives a detail description of our approach for mode decomposition, Section \ref{sec3} contains our experiments and results, some evaluations on characteristics of the proposed NMD, and comparisons with EMD and VMD on some given datas. Section \ref{sec4} conclude the contributions of this paper and discuss the future work.

\section{Methodology}\label{sec2}
\begin{figure}
  \centering
  \vspace{-1em}
  \includegraphics[width=2.5in]{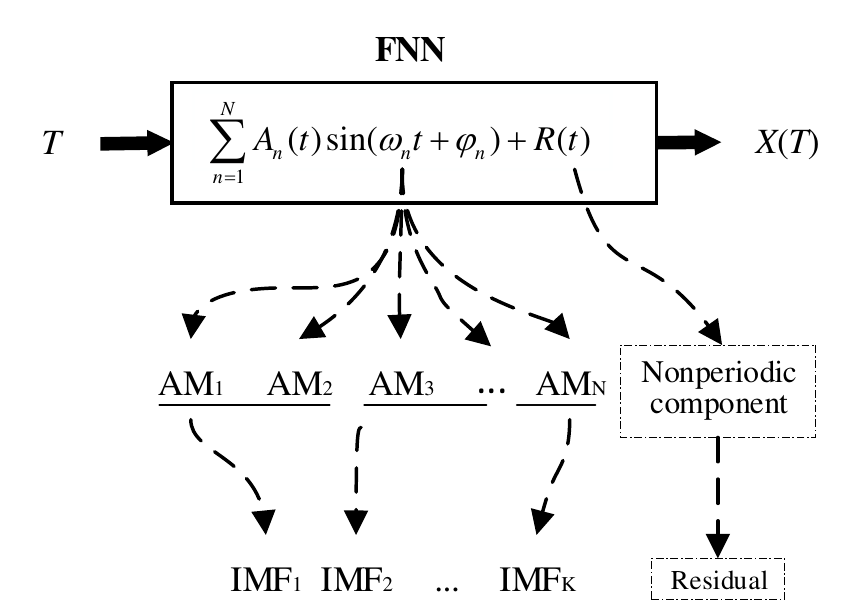}
  \vspace{-1em}
  \caption{Procedures of NMD.\label{Hu1}}
  \vspace{-1em}
\end{figure}

To improve the ability of collecting information of mode decomposition and study the decomposition algorithm with better adaptability and mathematical theory, 
this paper proposes a FNN based mode decomposition algorithm NMD. In NMD, firstly, the original data is expressed as the sum of amplitude-modulated (AM) signals and one non-periodic residual by means of Fourier neural network, and the corresponding energy of each AM signal are calculated. Secondly, a clustering algorithm is designed to divide the AM signals into several modes based on its energy. Finally, the AM signals in every single mode are summed to obtain the IMF components, constituting the final decomposition results with the residual in FNN. The frequency of obtained IMF are completely separated. Let $T$ and $X(T)$ represent the set of time sampling points of the data and the raw data to be decomposed, and the algorithm procedures is illustrated in Fig. \ref{Hu1}.
\subsection{Fourier Neural Network}
In IDFT, every discrete data $x(t)$ with length $N$ can be expressed as a linear combination of $N$ exponential functions,
\begin{equation}\label{eq2}
x(t)=\frac{1}{N}\sum_{n=0}^{N-1}X_{n}e^{j\frac{2\pi}{N}nt},0\leq t\leq N-1,
\end{equation}
where $X_{n}$ is the $n$th number returned by the DFT and $\frac{2\pi}{N}n$ is the $n$th term of the frequency. According to the periodicity of exponential function, the first $\lceil\frac{N}{2}\rceil-1$ ($\lceil\ \rceil$ represents a ceiling function) term starting from $n=1$ and the last $\lceil\frac{N}{2}\rceil-1$ term, of $N$ exponential functions in Eq. \ref{eq2} are conjugate symmetric. Letting $R$ be the term with $n=0$ of Eq. \ref{eq2} which is actually a constant, Eq. \ref{eq2} can be written as
\begin{equation}\label{eq3}
\begin{split}
x(t)=&\frac{1}{N}\sum_{n=1}^{\lceil\frac{N}{2}\rceil-1}\left[ (X_{n}-X_{N-n})j\sin(\frac{2\pi}{N}nt)\right.\\
&\left.+(X_{n}+X_{N-n})\sin(\frac{2\pi}{N}nt+\frac{\pi}{2})\right]+R \\
&+\left\{
     \begin{aligned}
     &X_{\frac{N}{2}}\sin(\pi t+\frac{\pi}{2}), N \, \rm{mod}\, 2=0 \\
     &0, N \, \rm{mod}\, 2=1
     \end{aligned}
     \right.,0\leq t\leq N-1.
\end{split}
\end{equation}

Time-varying signals can be viewed as a combination of one non-periodic trend and several periodic components. Based on this idea, the constant $R$ and the amplitude of sinusoidal waves are rewritten as functions, and we use a more uniform expression to describe Eq. \ref{eq3} as follows,
\begin{equation}\label{eq4}
x(t)=\frac{1}{N}\sum_{n=1}^{N-1}A_n(t)\sin(\omega_{n}t+\varphi_{n})+R(t).
\end{equation}

\begin{figure}
  \centering
  \includegraphics[width=3.2in]{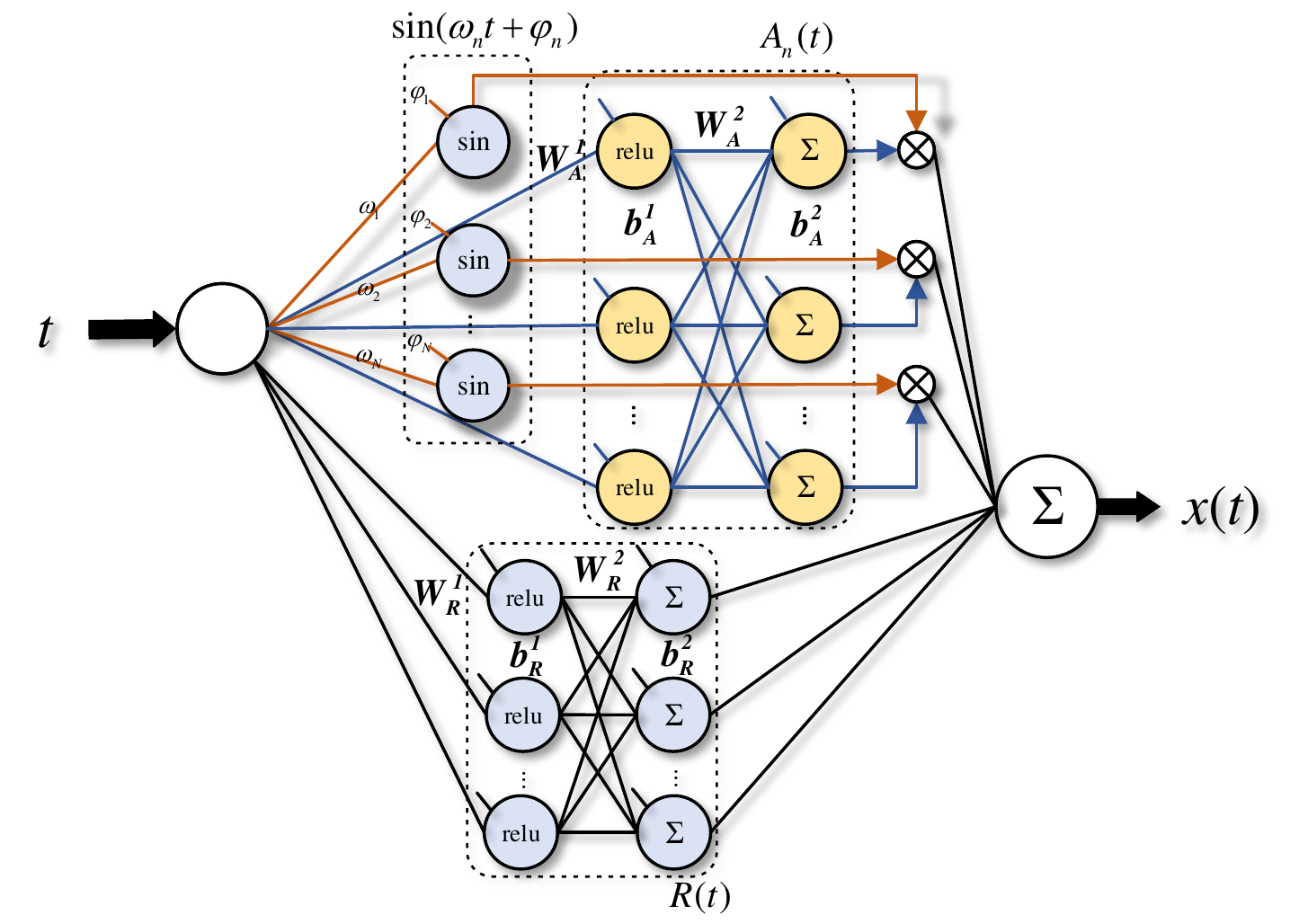}\\
  \vspace{-1em}
  \caption{Structure of FNN in NMD.}\label{Hu2}
  \vspace{-1em}
\end{figure}

Eq. \ref{eq4} meets the definition of IMF in Eq. \ref{eq2}.
We than design a FNN to fit the raw data in the form of Eq. \ref{eq4}, where $A_n(t)$, $\omega_{n}$, $b_{n}$ and $R(t)$ are parameters to be estimated. In FNN, we design two sub-networks to fit amplitude function $A_{n}(t)$ and frequency component $\sin(\omega_{n}t+\varphi_{n})$ of AM signals. Considering the sub-network for $A_{n}(t)$, a three layers neural networks with Relu active function in hidden layer is constructed. Considering the sub-network for $\sin(\omega_{n}t+\varphi_{n})$, a two layers neural networks with sinusoidal active function in output layer is constructed. We compute the dot-product of the outputs of these two sub-networks and obtain the periodic components of $x(t)$. Besides, we construct a three layers neural networks to fit the non-periodic component and add it to the periodic part. The structure of the constructed FNN is illustrated in Fig. \ref{Hu2}.

Our Fourier neural network model refers to the one in \cite{godfrey2017neural}, but what different from that is, the amplitude of every harmonic functions is developed from constant to time-varying function based on the different purpose of obtaining IMF in our research, which can reflect the relationship between time domain and frequency domain. The AM-FM-defined IMF can be broken into the sum of several AM signals. To simplify the neural network, we set the frequency parameters as a single parameter instead of a time-varying function in the frequency component. The constant frequency of every AM signal can help better control the separation of each IMF in the frequency domain.

Due to the strong periodicity of sinusoidal activate function, the loss function of FNN usually has strong periodicity and multiple local extreme points, which makes the parameters of FNN difficult to train \cite{gashler2016modeling}. Therefore, the initialization of parameters in FNN has a crucial impact for the training result of FNN. Based on Eq. \ref{eq3}, the $\omega_{n}$ is initialized as $2\pi\lfloor n/2\rfloor(0\le n\leq N)$, and the $\varphi_{n}$ is initialized as $(n$ mod $2)\times\pi/2$. This initialization ensures that the FNN can reconstruct the original data. Furthermore, after the training of neural network, the frequency and phase of AM signal can be fine tuned to be more consistent with the original data. For $A_{n}(t)$ and $R(t)$, parameters of them are initialized as small random values with uniform distribution, so that the neural network can learn the real periodic and non-periodic characteristics in the data without being affected by initialization.

In the neural network constructed in this paper, there is a trend that non-periodic parts are fitted by neurons of periodic parts \cite{gashler2016modeling}, so it is necessary to add a regularization term on the parameters of amplitude function to prevent it. Similarly, the regularization can make the parameters of periodic part concentrate near the main frequency components. $L^1$ regularization can make the parameter distribution be more sparse, and $L^2$ regularization make the parameter distribution be more average. According to the purpose of this paper, we use $L^1$ regularization in our work.

We obtain the frequency energy $\overline{A_{n}}=\frac{1}{N}\sum^{N}_{t=1}A_{n}^{2}(t)$ from every $A_{n}(t)$ of the trained FNN, which can be viewed as the weight of the corresponding frequency in the original data. The distribution of AM signals into IMFs will be based on this energy.

\subsection{Energy-weighted Frequency Clustering}
We design a clustering algorithm to divide all AM signals obtained from trained FNN into modes. First of all, we choose the primary frequency components from these AM signals according to corresponding $\overline{A_{n}}$. Let $P(0<P<1)$ represents a energy threshold, rank $\overline{A_{n}}$ from height to low and return top minimum $M$ entities whose sum of energy makes up more than $P$ as primary frequency components.

After obtaining primary frequency components, the number and frequency boundaries of modes are obtained based on them. In the beginning, we assume that each principal frequency component represents a principal component of a different mode. Taking the frequency of all AM signals as the x-axis and the corresponding $\overline{A_{n}}$ as the y-axis, a frequency-energy model are established. The points of all primary frequency components are projected onto the x-axis, and the energy $\overline{A_{n}}$ of each point are taken as weight. For the $i$th initial principal component point, draw a circle $C_{i}$ with radius $r$, and calculate the sum of weights of all primary frequency components falling in $C_{i}$, $\sum_{j\in{C_{i}}}\overline{A_{j}}$, as the weight of this mode. Define the similarity $S(i,i+1)$ and $S(i+1,i)$ between $C_{i}$ and $C_{i+1}$ as the proportions of overlapping part's weight in two adjacent circle $C_{i}$ and $C_{i+1}$. $S(i,i+1)$ and $S(i+1,i)$ are expressed as
\begin{equation}\label{eq5}
S(i,i+1)=\frac{\sum_{j\in(C_{i}\cap C_{i+1})\overline{A_{j}}}}{\sum_{j\in{C_{i}}}\overline{A_{j}}},
S(i+1,i)=\frac{\sum_{j\in(C_{i}\cap C_{i+1})\overline{A_{j}}}}{\sum_{j\in{C_{i+1}}}\overline{A_{j}}}
\end{equation}

When any of $S(i,i+1)$ and $S(i+1,i)$ exceeds $80\%$, these two initial principal components of different modes can be viewed belonging to a same mode, so combine the $i$th and $i+1$th primary frequency components into one mode and update the weight in this mode, then measure the similarity between the updated mode and the next adjacent initial principal component. Otherwise, the $i$th and $i+1$th primary frequency components are considered to be in two different modes, than measure the similarity between the $i+1$th and $i+2$th initial principal components. The algorithm begin from the first primary frequency component, and end after all $M$ primary frequency components have been processed.

\begin{figure}
  \centering
  \includegraphics[width=3.2in]{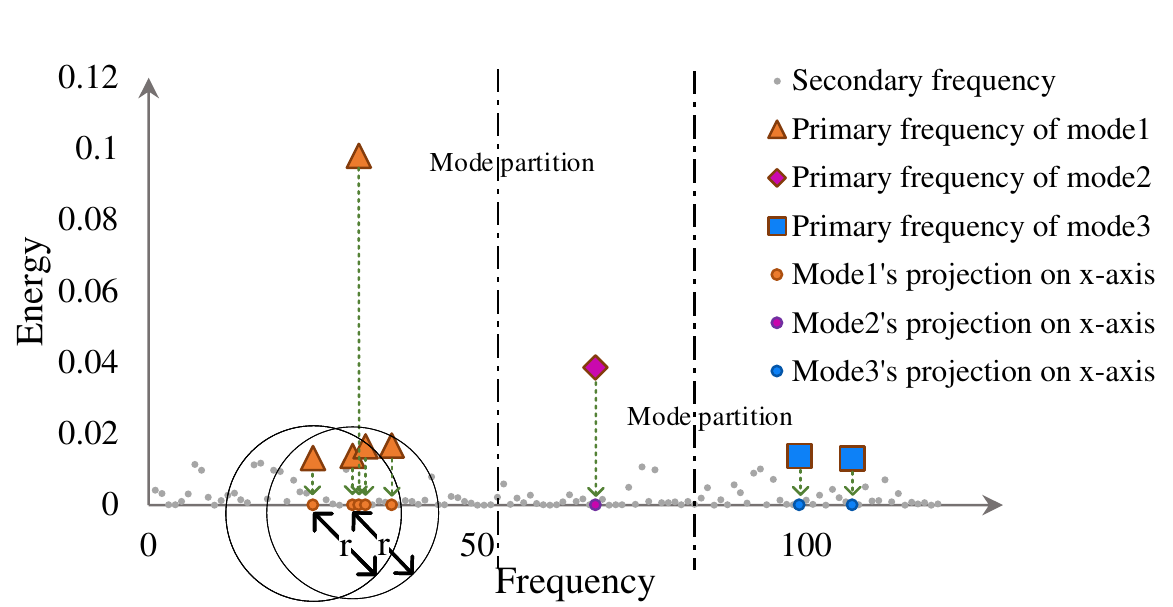}\\
  \vspace{-1em}
  \caption{Energy based mode division.}\label{Hu3}
  \vspace{-1em}
\end{figure}

After distributing all the primary frequency components into modes, the midpoint of the adjacent primary frequency components between two adjacent different modes is taken as the mode division point. In the set of the first, the $N$th and all mode division points, sum all primary frequency components and other AM signals between each two adjacent points to obtain the final IMF. The diagram of energy based frequency clustering is shown in \ref{Hu3}.
\section{NMD characteristic analysis and experimental results}\label{sec3}
In this section we illustrate the characteristics and effectiveness of proposed NMD on several examples. First, we investigate the effects of varied regularization coefficient $\lambda$ and energy threshold $P$ to NMD results. Second, we apply EMD, VMD and NMD to 4 test datas, compare and analyze these decomposition results from the perspective of the ability of collecting features, orthogonality, completeness, et. al. Finally, we shift our attention to the FNN's potential capability of extrapolating and predicting data and investigate the feature mining from the perspective of extrapolation.


We use Pytorch's implementation of neural networks in the experiments. We utilize stochastic gradient descent algorithm with Nesterov momentum to train the neural network, and choose the mean square error (MSE) between the output of FNN and real data as the loss function. We normalize the time related to each sample and the sampling data so that all training are between 0 and 1 on the x-axis and y-axis.
\subsection{Hyper-parameters of NMD}\label{sec3.1}

The NMD proposed in this paper has two important hyper-parameters, regularization coefficient $\lambda$ in $L^{1}$ regularization of amplitude sub-network in FNN and the energy threshold $P$ in frequency clustering. $\lambda$ would highly affect the distribution of energy $\overline{A_{n}}$, and $P$ would affect the primary frequency components selected from AM signals, further affecting the number of modes. To study the influence of hyper-parameters on decomposition results, we select two synthetic datas with multi-frequency components. The impact of noise on NMD results is also presented.
\subsubsection{Regularization Coefficient $\lambda$}

We first analyze the impact of regularization coefficient $\lambda$ on decomposition results of $x_{\lambda}(t)$. $x_{\lambda}(t)$ composed of a quadratic trend, two different harmonics, a chirp signal and a Gaussian noise of $N(0,0.1)$. $\lambda$ start from 0 and increase with step size 0.001. We choose the decomposition results which change significantly in this process to present.
\begin{equation}\label{eq5}
x_{\lambda}(t)=4t^{2}+\cos(8\pi t)+2\cos(40\pi t)+\cos(50\pi t+20\pi t^{2})+\eta
\end{equation}

\begin{figure}
		\begin{subfigure}[t]{0.18\textwidth}
            \centering
			\includegraphics[width=\textwidth]{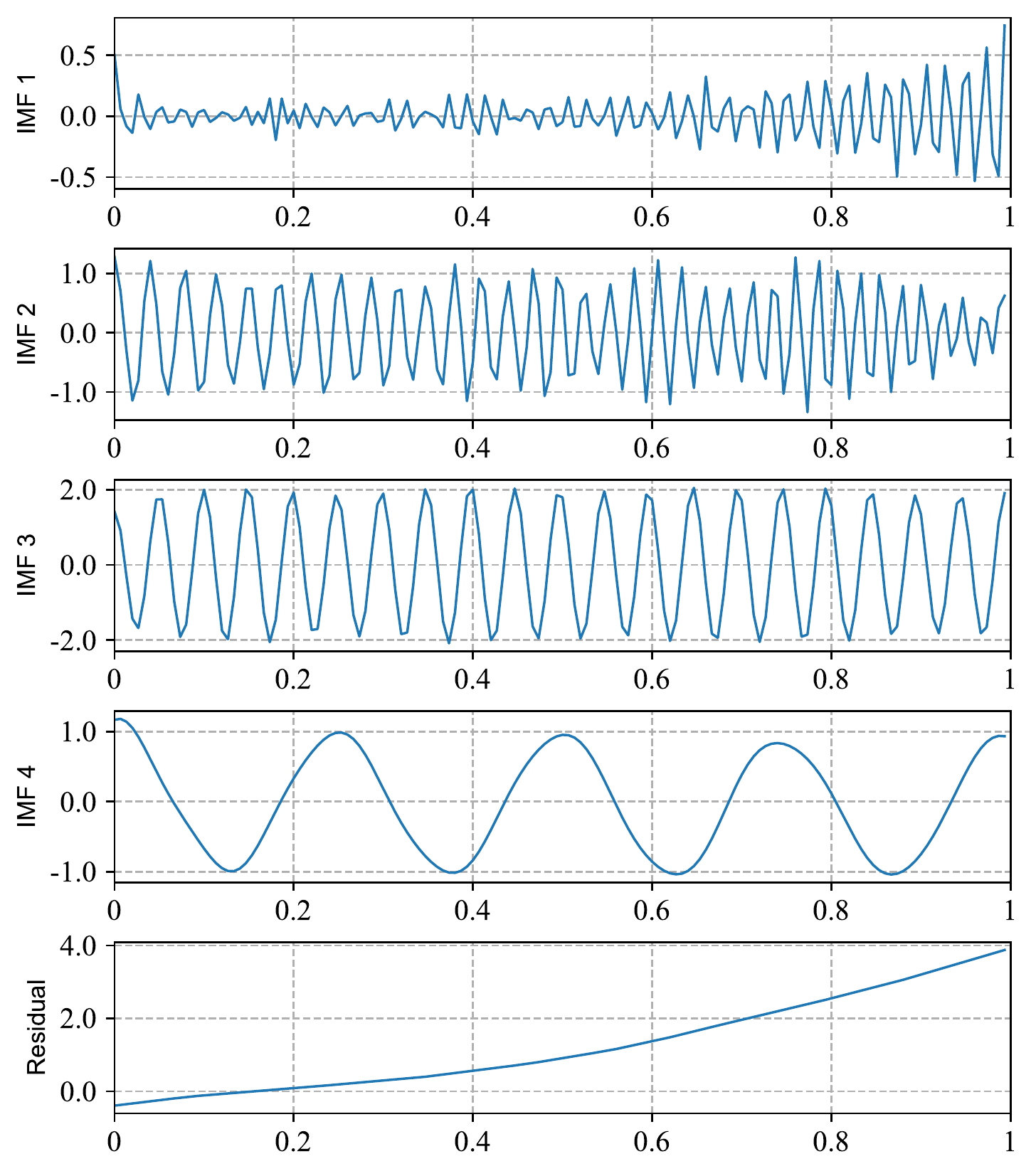}
            \label{Hu4a}
            \vspace{-2em}
			\caption{$\lambda$=0.001}
		\end{subfigure}
		\begin{subfigure}[t]{0.18\textwidth}
			\centering
			\includegraphics[width=\textwidth]{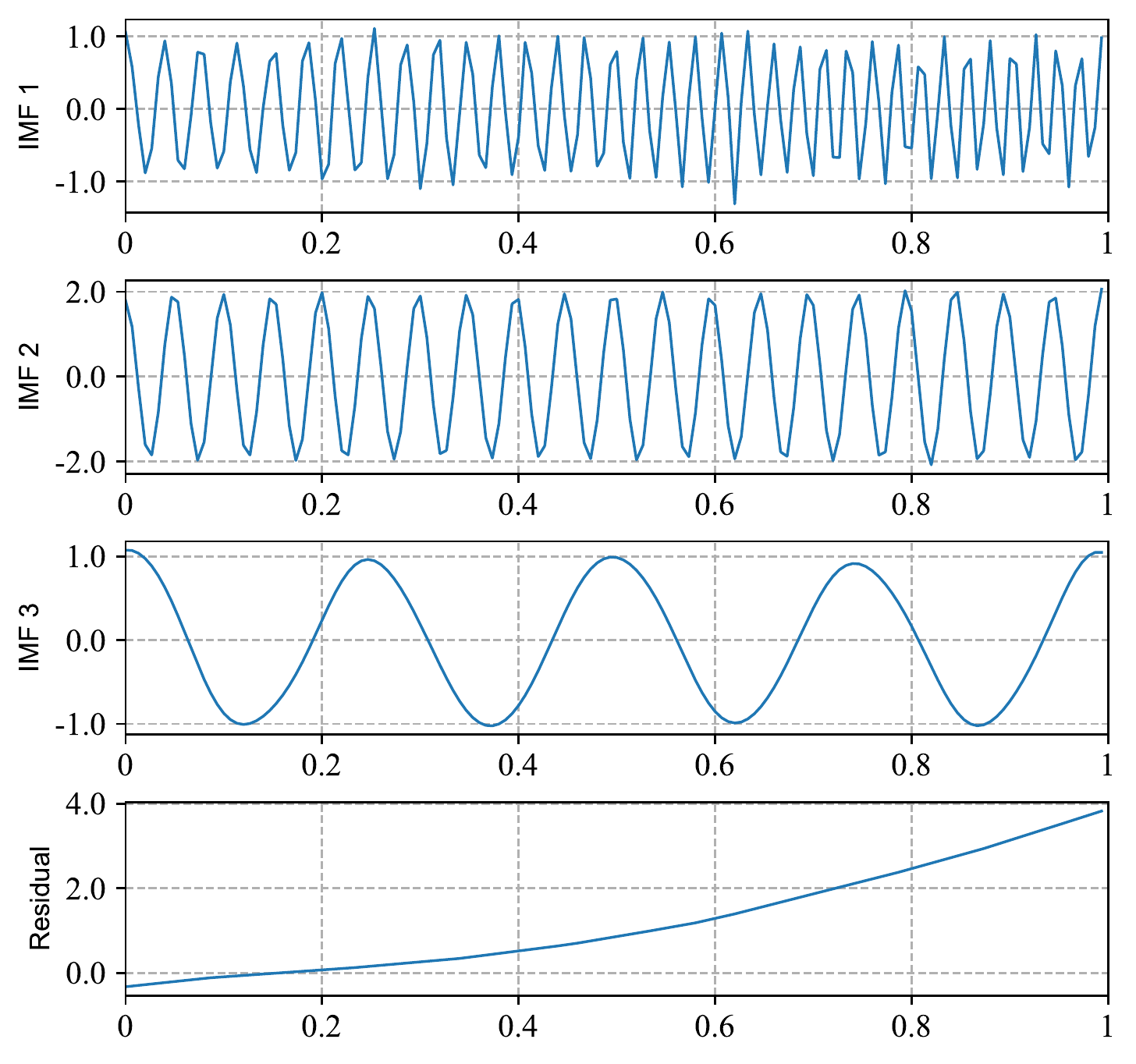}
            \label{Hu4b}
            \vspace{-2em}
			\caption{$\lambda$=0.1}
		\end{subfigure}
		\begin{subfigure}[t]{0.18\textwidth}
			\centering
			\includegraphics[width=\textwidth]{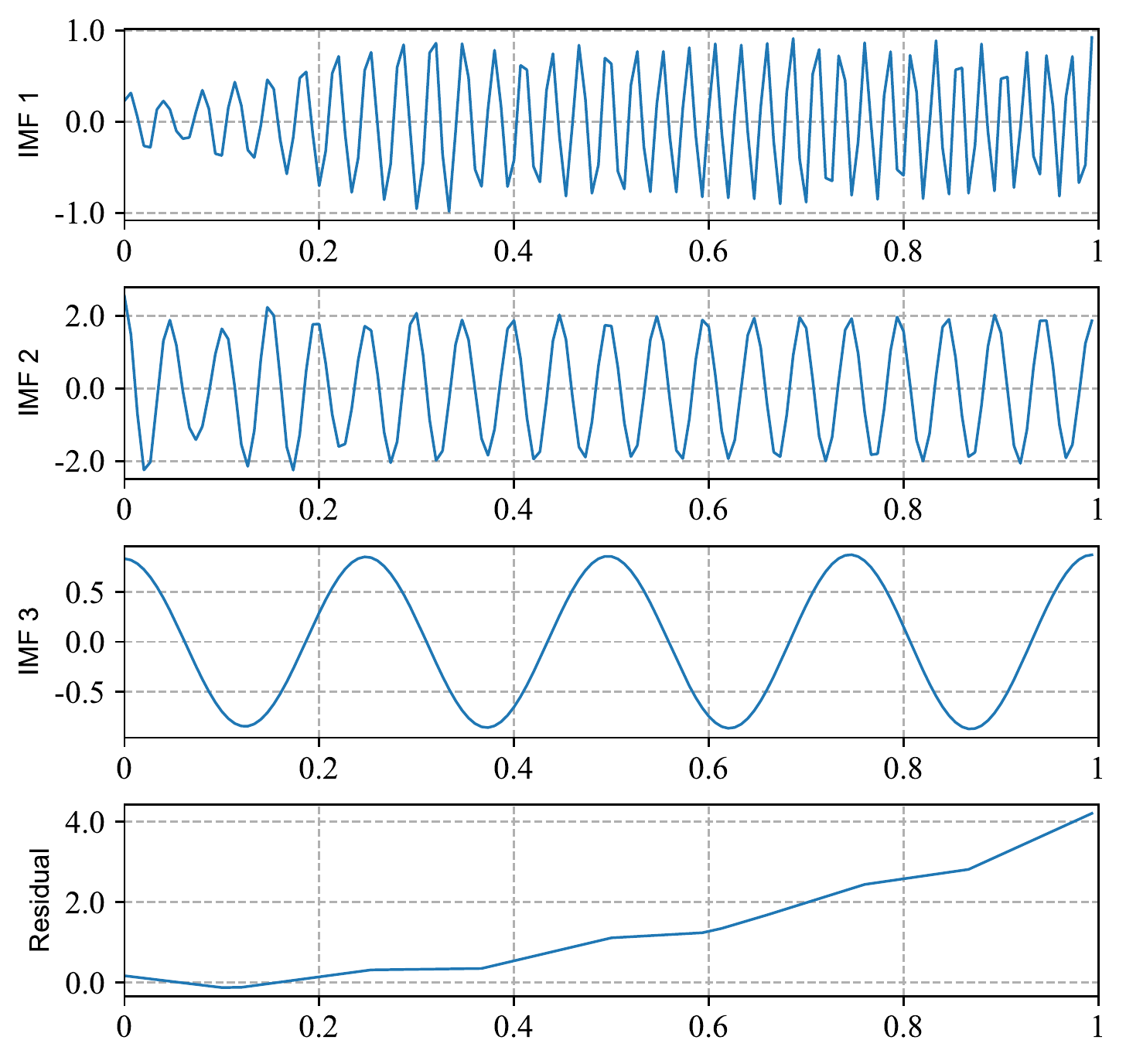}
            \label{Hu4c}
            \vspace{-2em}
			\caption{$\lambda$=0.2}
		\end{subfigure}
		\begin{subfigure}[t]{0.18\textwidth}
			\centering
			\includegraphics[width=\textwidth]{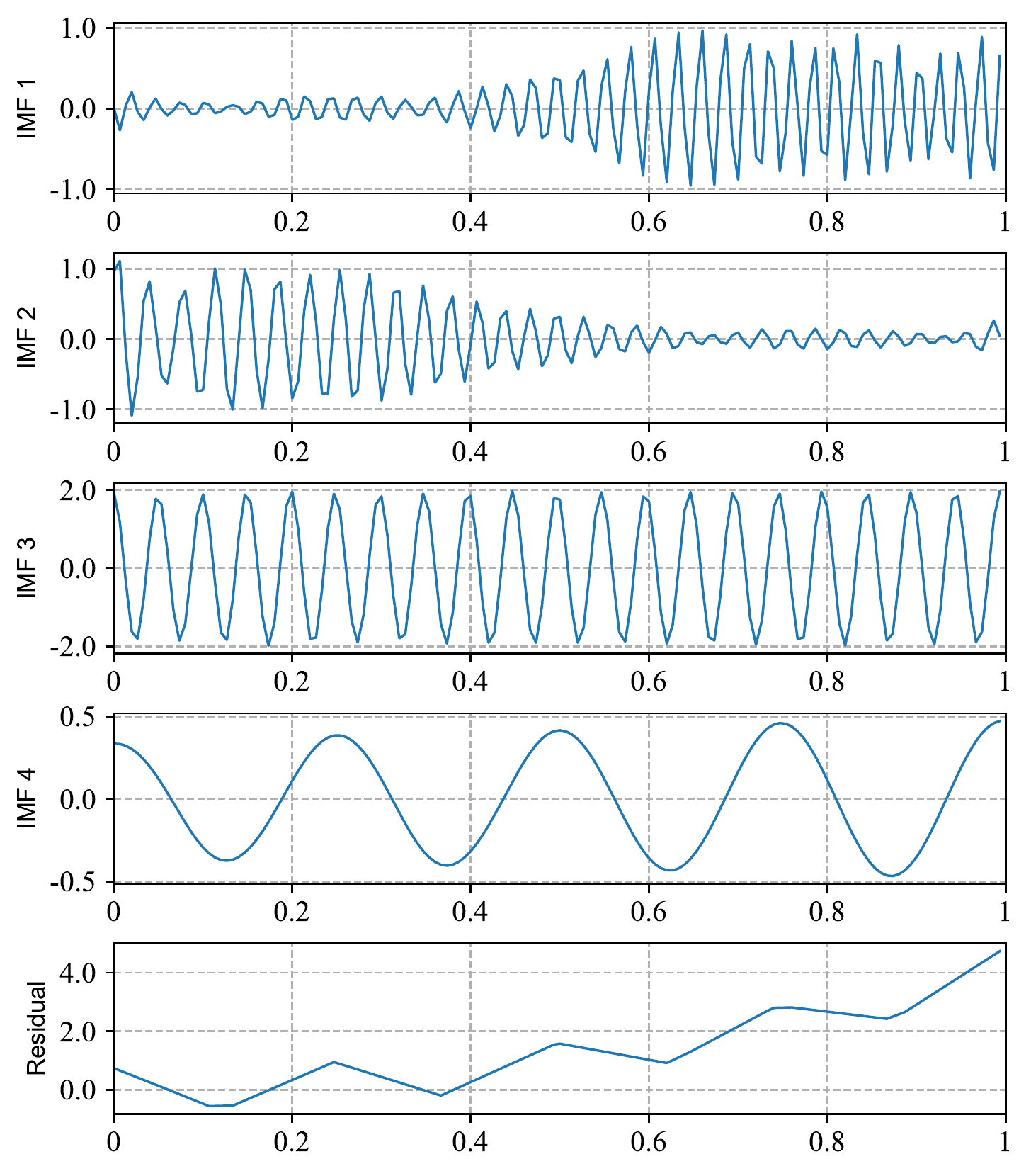}
            \label{Hu4d}
            \vspace{-2em}
			\caption{$\lambda$=0.5}
		\end{subfigure}
        \vspace{-1em}
        \caption{NMD's decomposition results on $x_{\lambda}(t)$. (a)-(d) Decomposed components with $\lambda=0.001,0.1,0.2,0.5$ respectively.}
        \label{Hu4}
	\end{figure}

\begin{figure}
		\begin{subfigure}[t]{0.18\textwidth}
			\centering
			\includegraphics[width=\textwidth]{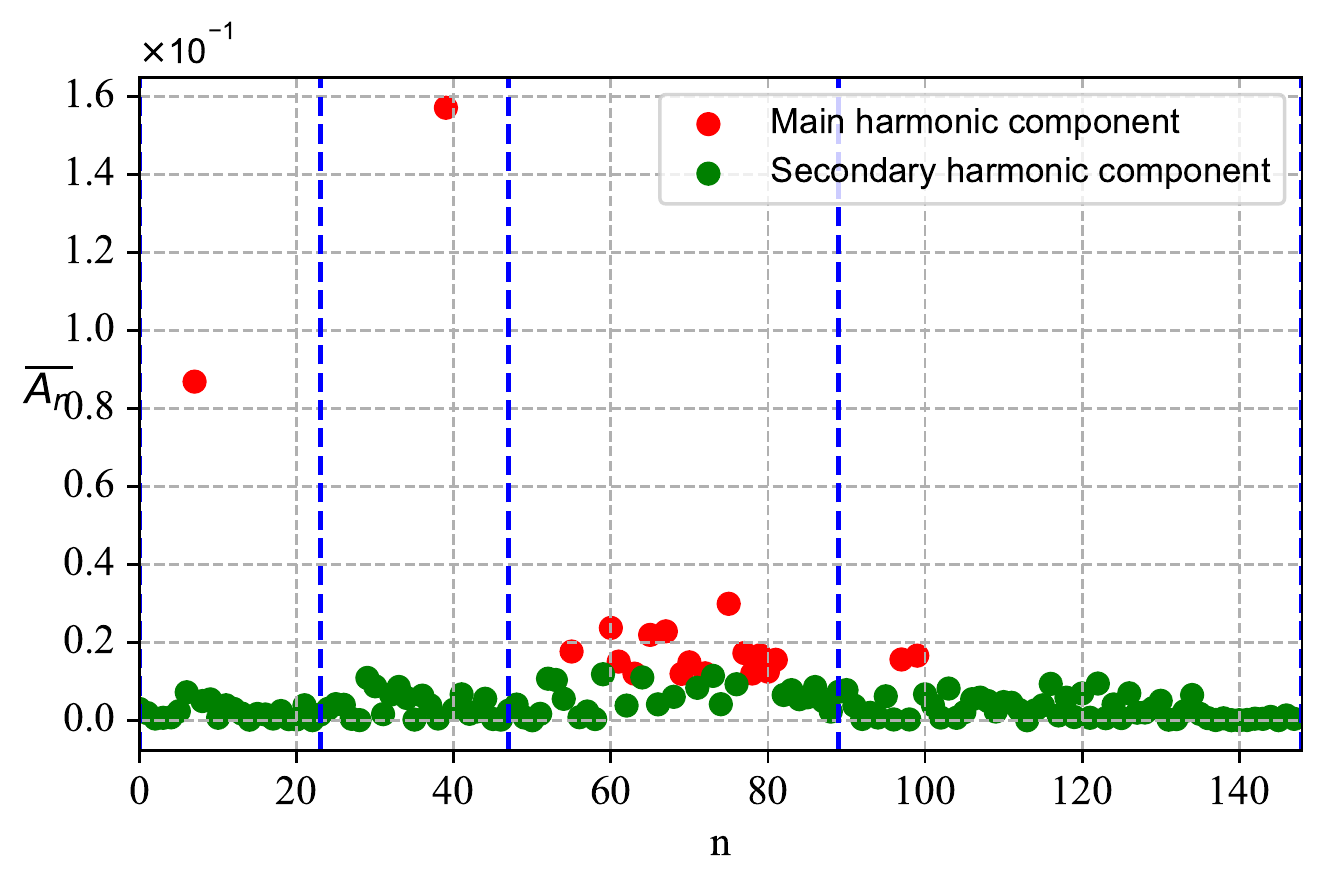}
            \label{Hu5a}
            \vspace{-2em}
			\caption{$\lambda$=0.001}
		\end{subfigure}
		\begin{subfigure}[t]{.18\textwidth}
			\centering
			\includegraphics[width=\textwidth]{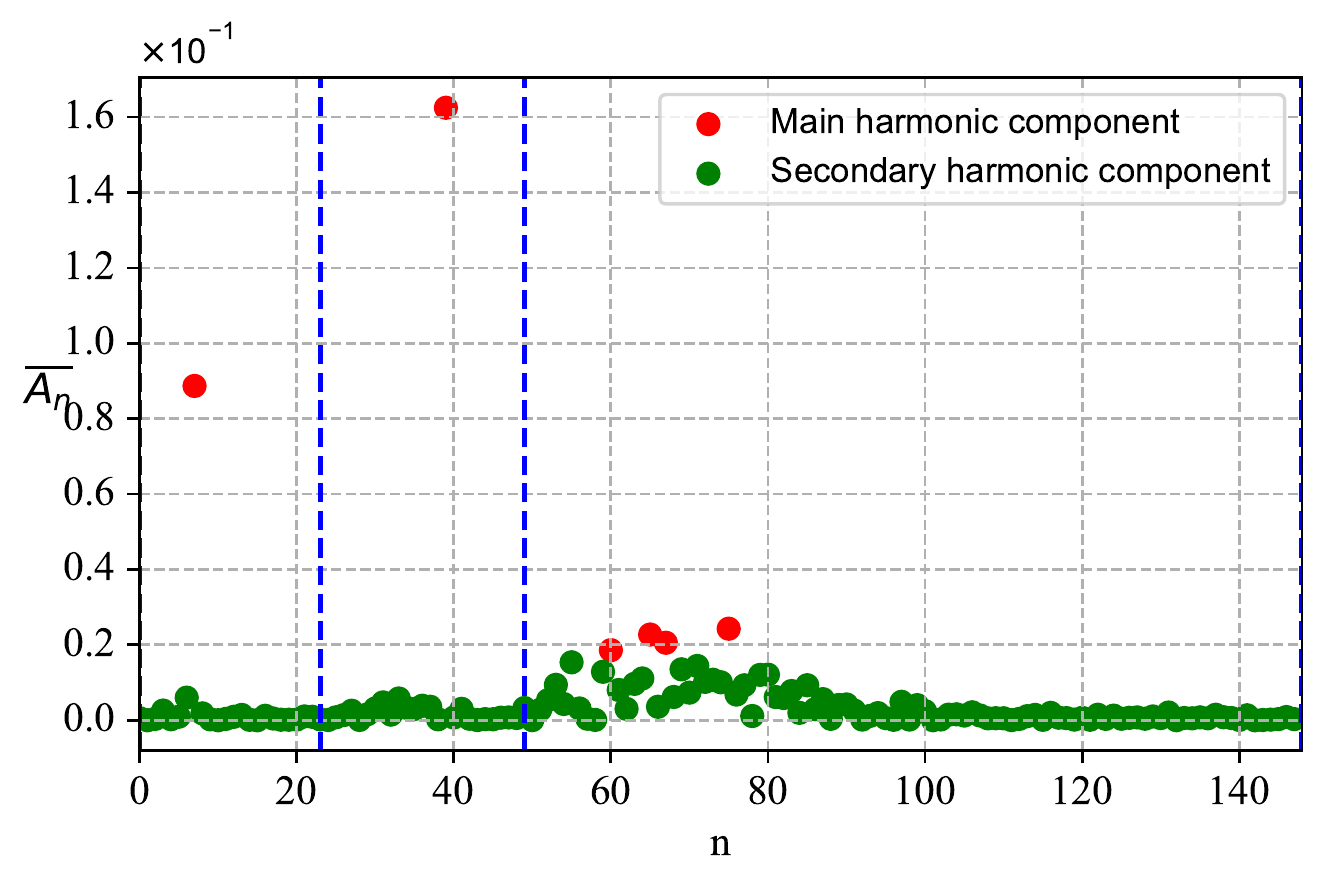}
            \label{Hu5b}
            \vspace{-2em}
			\caption{$\lambda$=0.1}
		\end{subfigure}
		\begin{subfigure}[t]{.18\textwidth}
			\centering
			\includegraphics[width=\textwidth]{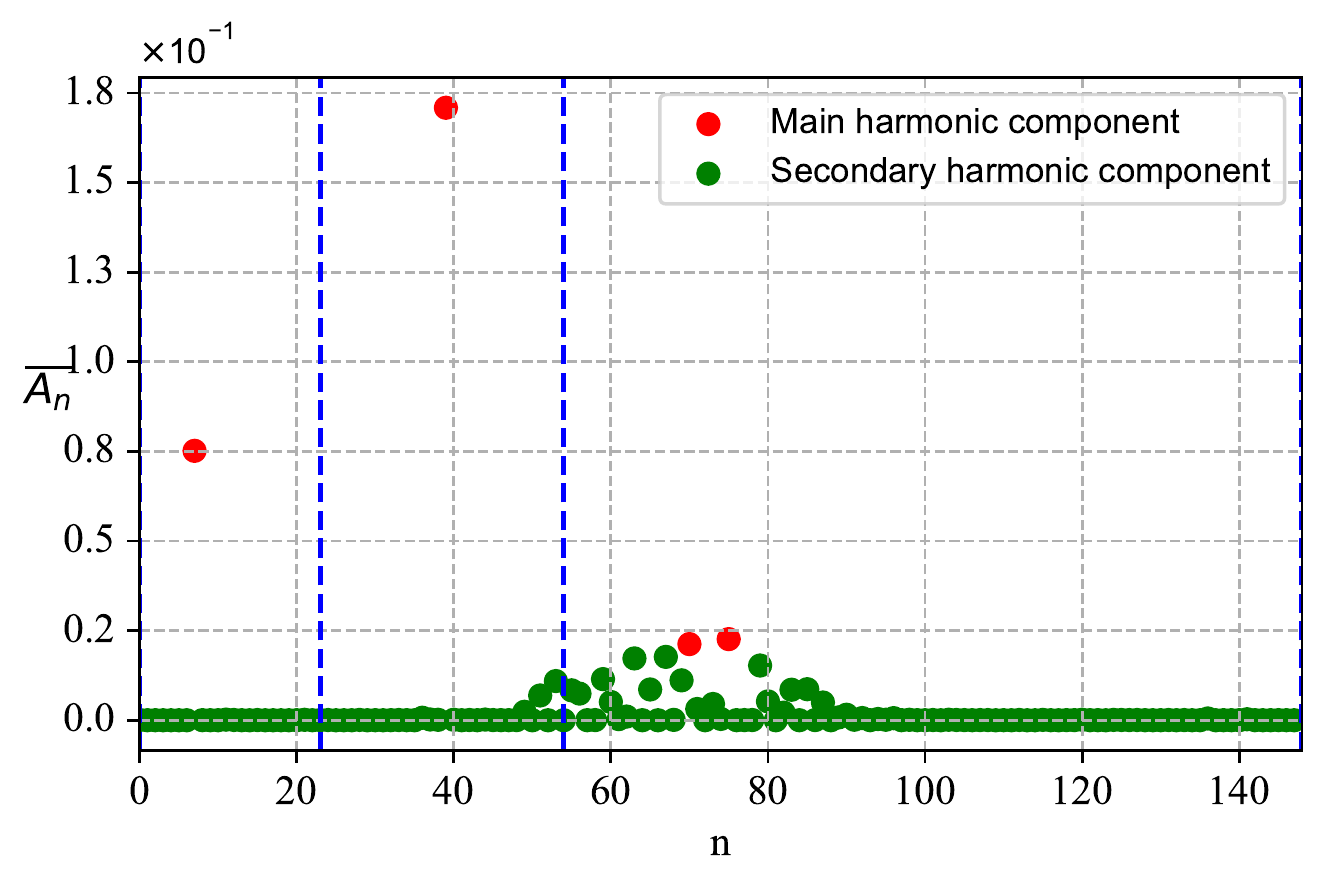}
            \label{Hu5c}
            \vspace{-2em}
			\caption{$\lambda$=0.2}
		\end{subfigure}
		\begin{subfigure}[t]{.18\textwidth}
			\centering
			\includegraphics[width=\textwidth]{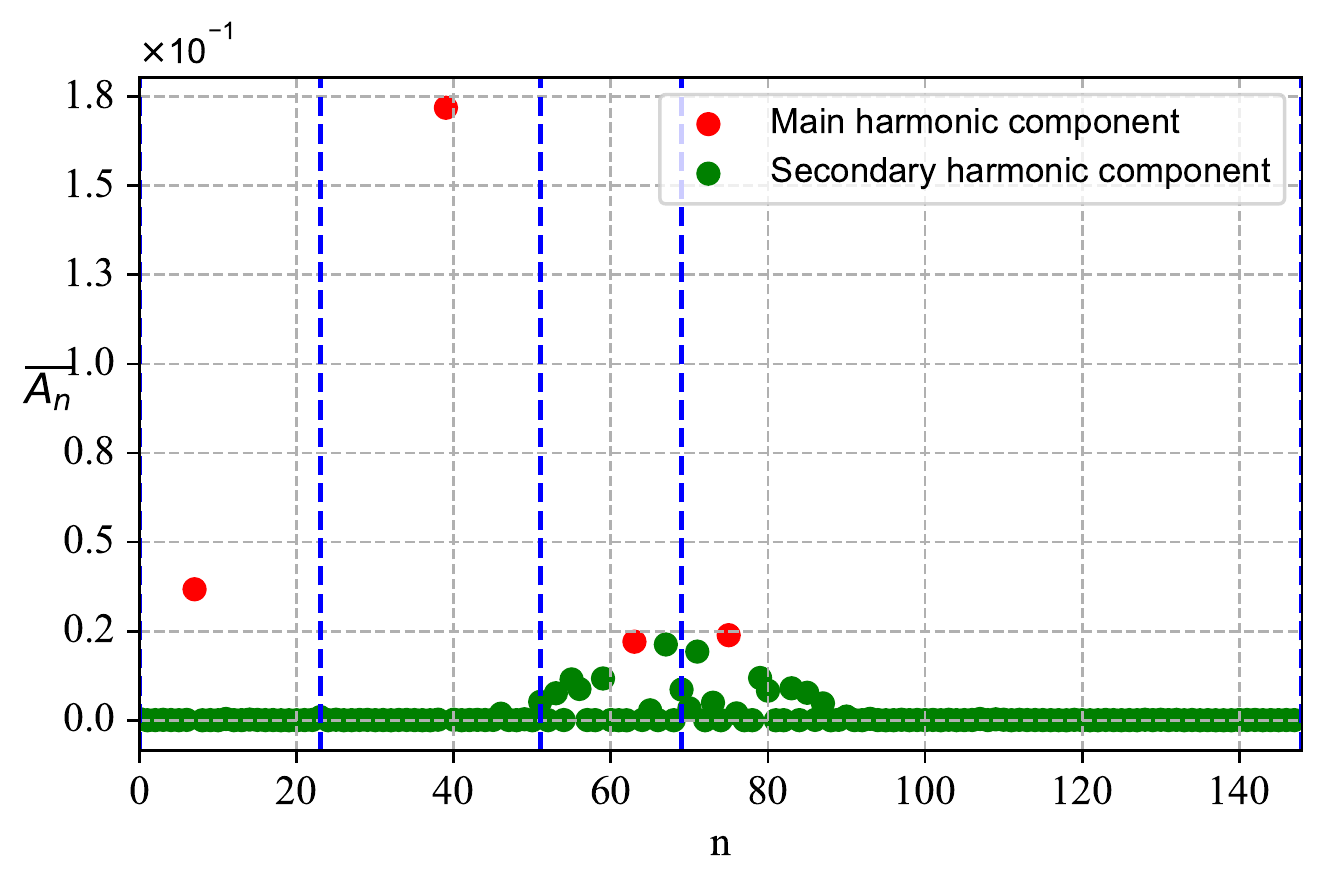}
            \label{Hu5d}
            \vspace{-2em}
			\caption{$\lambda$=0.5}
		\end{subfigure}
        \vspace{-1em}
        \caption{Energy distribution of AM signals of $x_{\lambda}(t)$ trained by FNN. (a)-(d) Distributions with $\lambda=0.001,0.1,0.2,0.5$ respectively.}
        \vspace{-1em}
        \label{Hu5}
	\end{figure}

As $\lambda$ grows from 0 to 0.5, NMD shows us four distinctly different mode decomposition results, whose details and energy distribution $\overline{A_{n}}$ are illustrated in Fig. \ref{Hu4} and Fig. \ref{Hu5}.

Fig. \ref{Hu4} shows that $\lambda$ is not directly related to the number of decomposed components, but influence the distribution of $\overline{A_{n}}$ to affect decomposition results. The last two IMFs of 4 decomposition results always manage to recover frequency 8 and frequency 20 of original data respectively. When $\lambda=0.001$, the distribution of $\overline{A_{n}}$ is quite dispersed so we obtain the most primary frequency components, and the chirp component in original data is decomposed into IMF1 and IMF2 according to the clustering on the dispersed distribution. When $\lambda$ increase to 0.01, the energy distribution becomes concentrated, so the clustering algorithm clusters the frequency energy of chirp component in one mode. As $\lambda$ increasing to 0.2, most of the frequency energy are driven to 0, and the energy of chirp component distribute more uniform and gentler, so that some frequency energy of chirp component leak into the mode of frequency 8, resulting distorted recovering of original data components. Finally, when $\lambda$ reach 0.5, the continuously gentle becoming chirp component energy make this component be divided into two modes once again, and the residual shows a periodical fluctuate indicating that the neurons of non-periodic $R(t)$ model some periodic components of original data, which is not what our needs. Hence, the experiments is finished. Although $x_{\lambda}(t)$ has some noise, it has little effect on the mode decomposition results because of the small variance.

In conclusion, in practical application of NMD, we can adjust the regularization coefficient according to the actual requirements to control the sparsity of energy distribution, and obtain the required results in different scenarios. Inappropriate regularization coefficient may distort the decomposition results.

\subsubsection{Energy Therhold $P$}

Refering to the noise test data in \cite{dragomiretskiy2013variational}, $x_{P}(t)$ is selected to study the impact of $P$ on decomposition results and noise immunity. $x_{P}(t)$ is composed of three pure harmonics and a Gaussian noise with standard deviation 0.1. The 0.1 standard deviation of noise indicates that the noise takes up a larger constituent in $x_{P}(t)$ than the highest and least frequency composition.
\begin{equation}\label{eq5}
x_{P}(t)=\cos(4\pi t)+\frac{1}{4}\cos(48\pi t)+\frac{1}{16}\cos(200\pi t)+\eta
\end{equation}

\begin{figure}
		\begin{subfigure}[t]{0.15\textwidth}
            \centering
			\includegraphics[width=\textwidth]{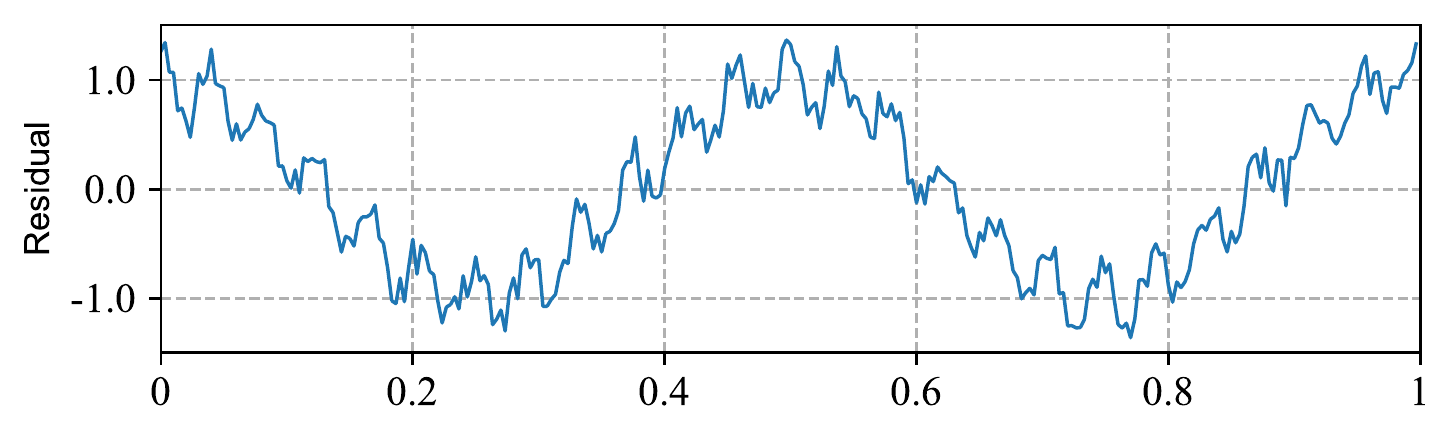}
            \label{Hu6a}
            \vspace{-2em}
			\caption{$P$=90\%}
		\end{subfigure}
		\begin{subfigure}[t]{0.15\textwidth}
			\centering
			\includegraphics[width=\textwidth]{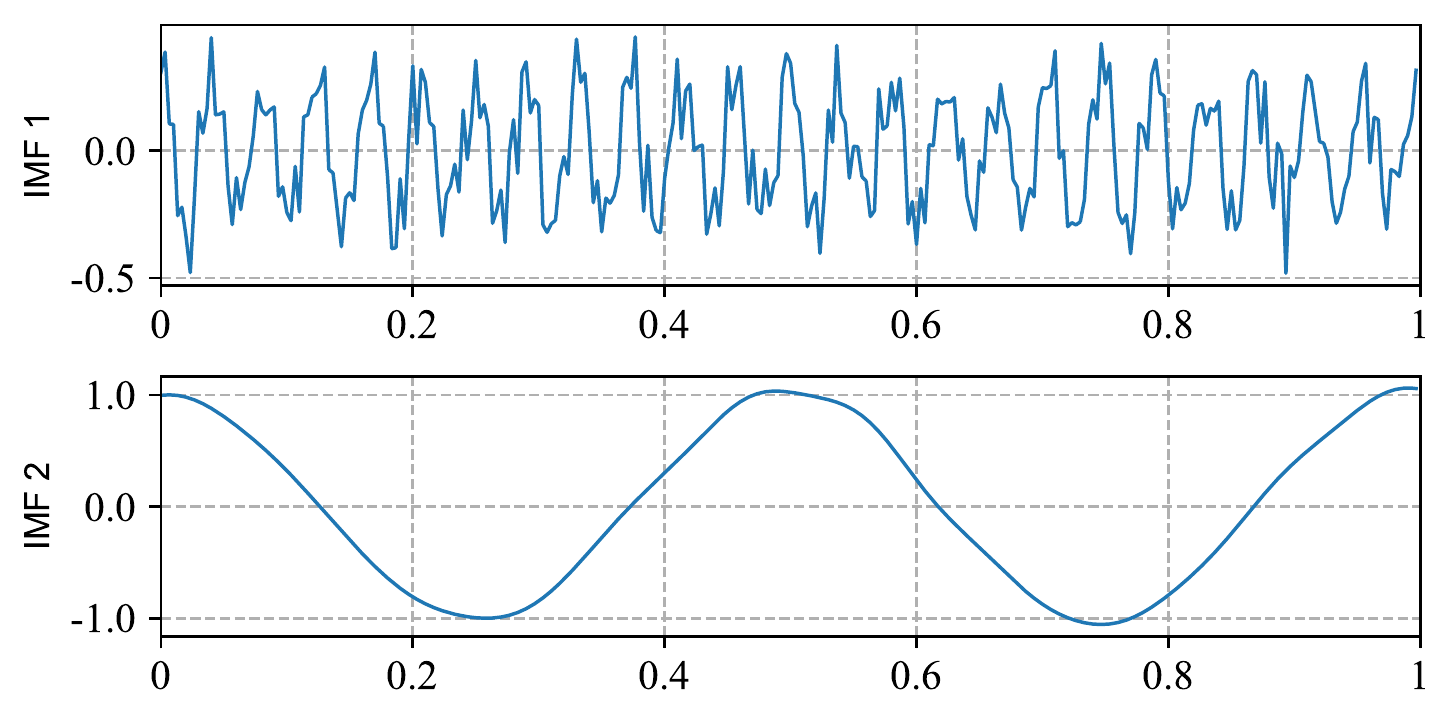}
            \label{Hu6b}
            \vspace{-2em}
			\caption{$P$=92\%}
		\end{subfigure}
		\begin{subfigure}[t]{0.15\textwidth}
			\centering
			\includegraphics[width=\textwidth]{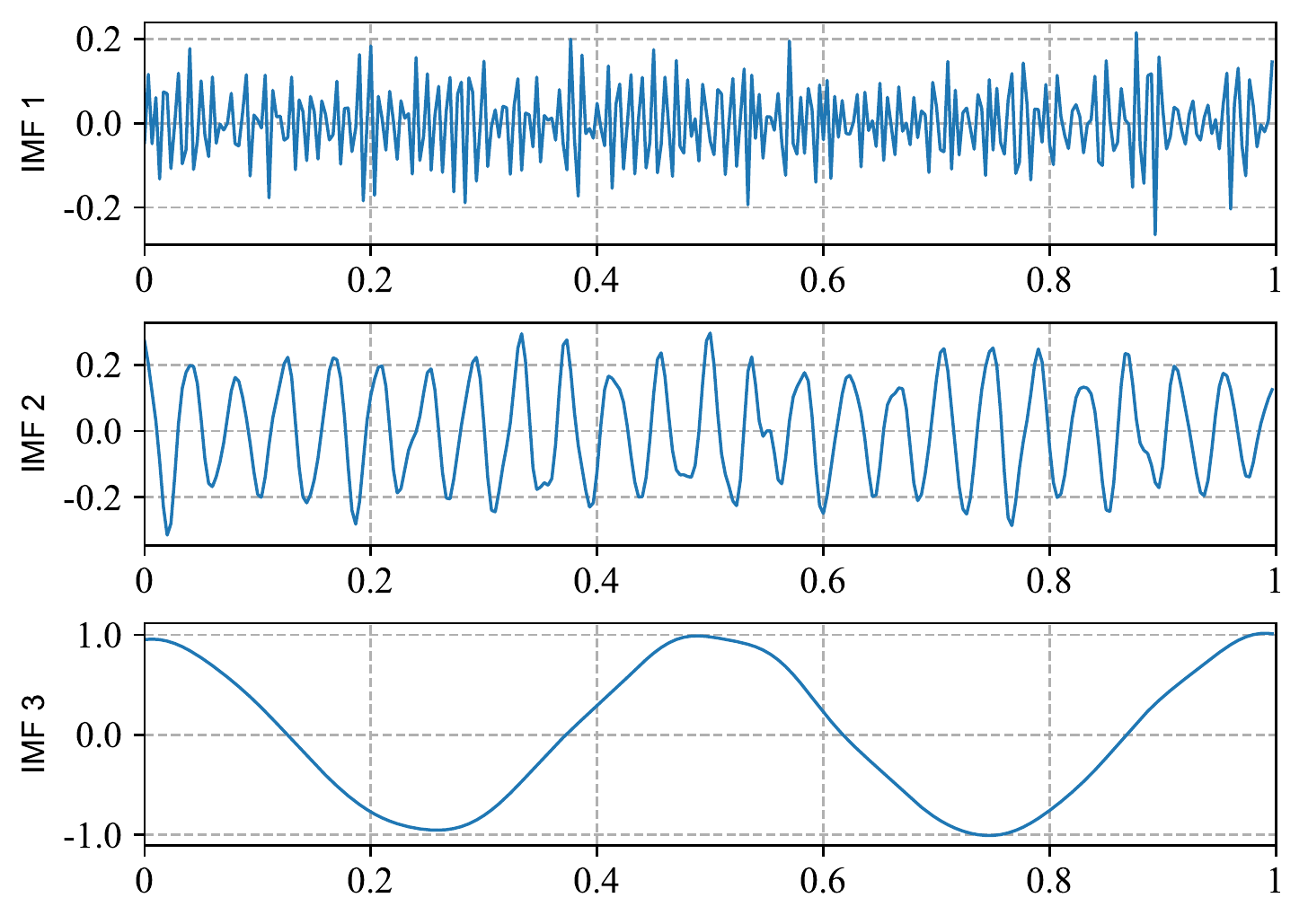}
            \label{Hu6c}
            \vspace{-2em}
			\caption{$P$=98\%}
		\end{subfigure}
        \vspace{-1em}
        \caption{NMD's decomposition results on $x_{P}(t)$. (a)-(c) Decomposed components with $P=90\%,92\%,98\%$ respectively.}
        \vspace{-1em}
        \label{Hu6}
	\end{figure}

\begin{figure}
		\begin{subfigure}[t]{0.15\textwidth}
			\centering
			\includegraphics[width=\textwidth]{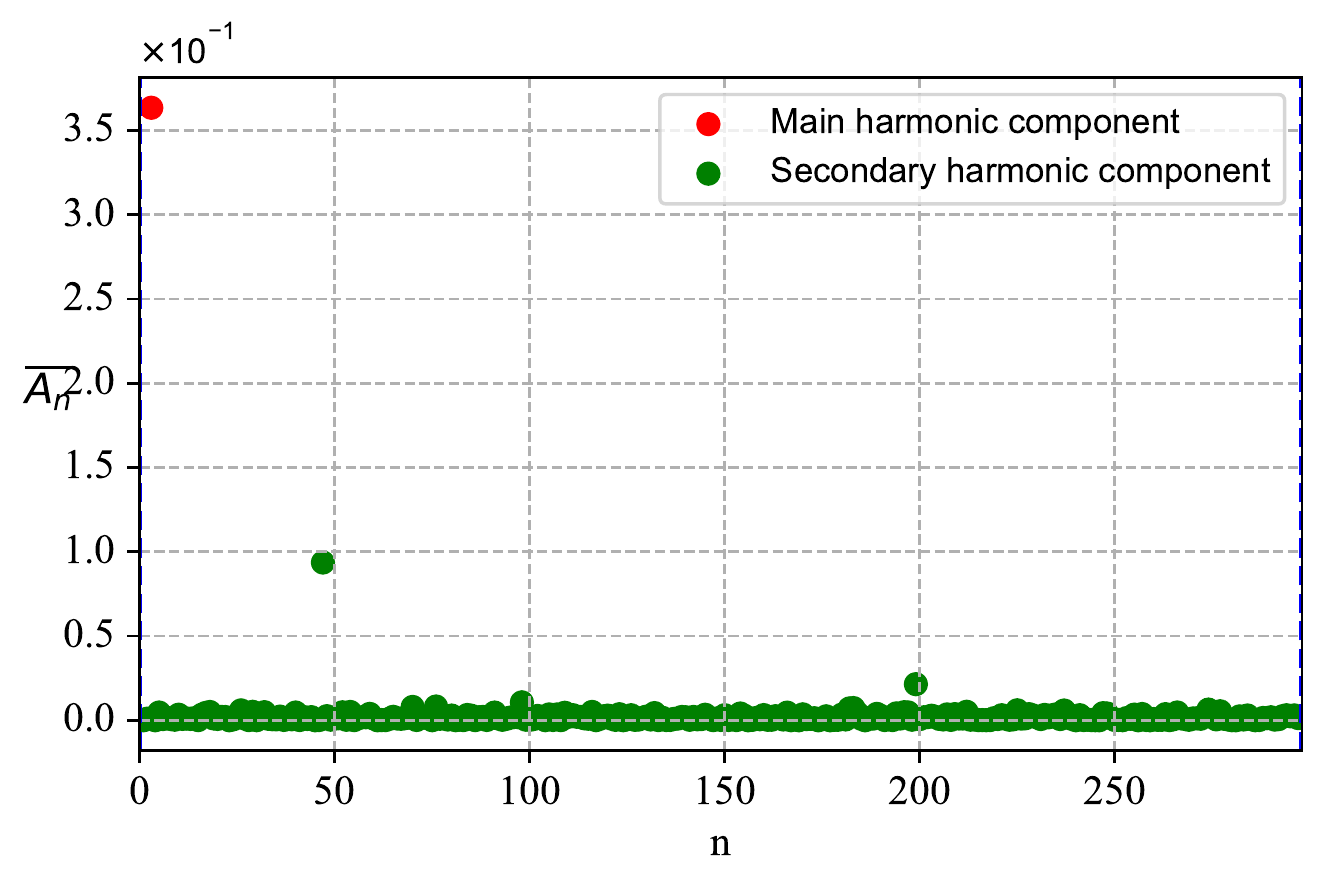}
            \label{Hu7a}
            \vspace{-2em}
			\caption{$P$=90\%}
		\end{subfigure}
		\begin{subfigure}[t]{.15\textwidth}
			\centering
			\includegraphics[width=\textwidth]{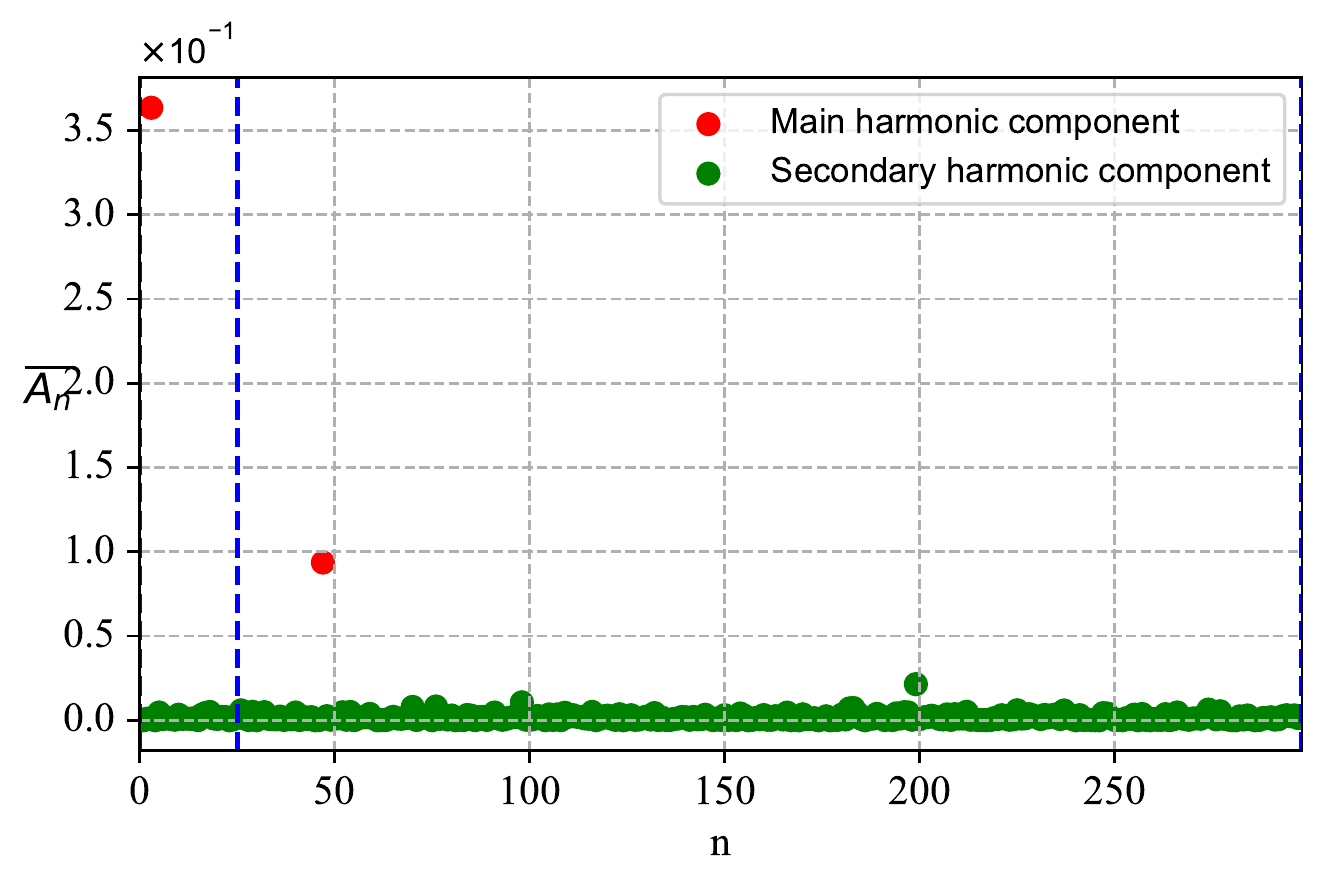}
            \label{Hu7b}
            \vspace{-2em}
			\caption{$P$=92\%}
		\end{subfigure}
		\begin{subfigure}[t]{.15\textwidth}
			\centering
			\includegraphics[width=\textwidth]{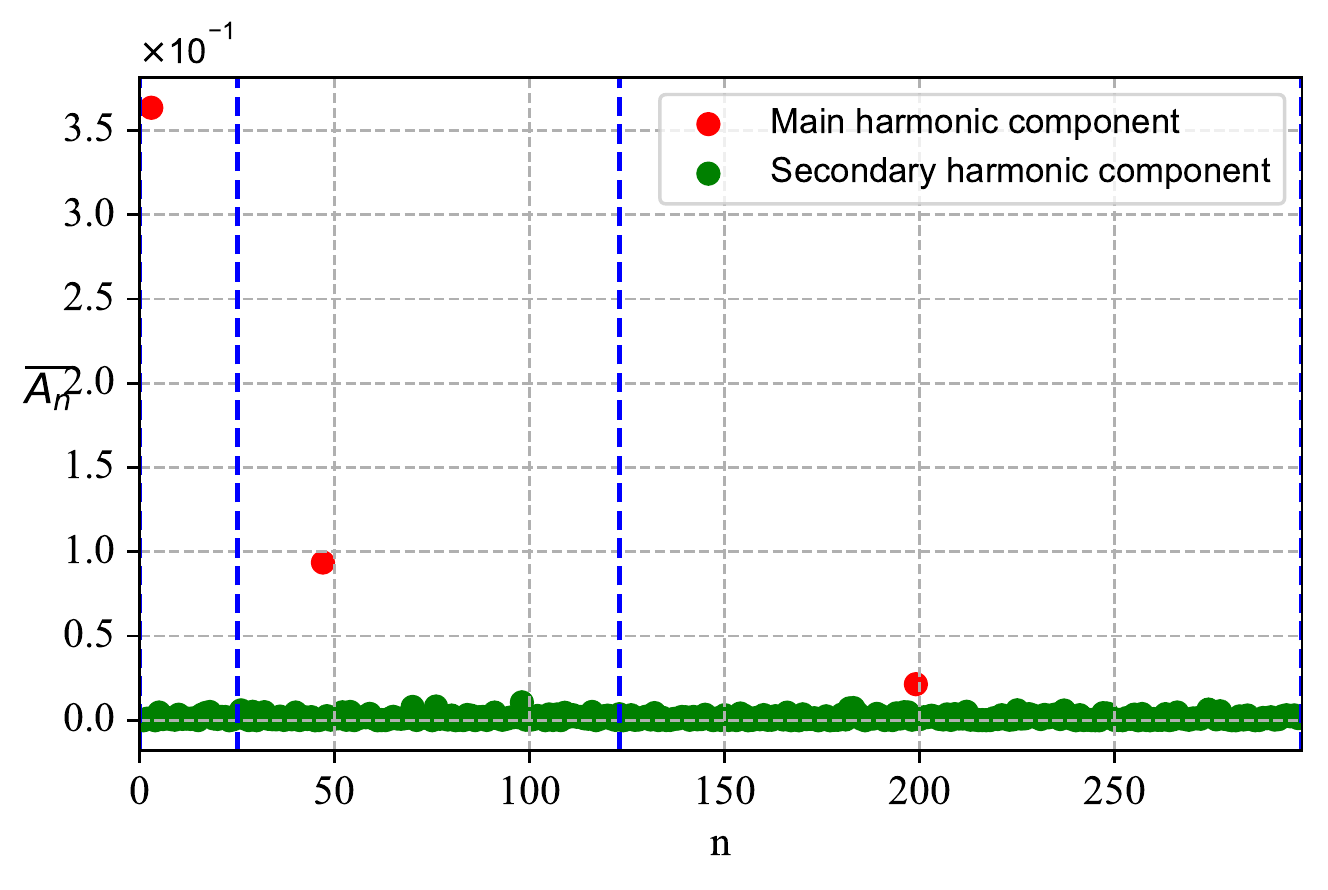}
            \label{Hu7c}
            \vspace{-2em}
			\caption{$P$=98\%}
		\end{subfigure}
        \vspace{-1em}
        \caption{Energy distribution of AM signals of $x_P(t)$ trained by FNN. (a)-(c) Distributions with $P=90\%,92\%,98\%$ respectively.}
        \vspace{-1em}
        \label{Hu7}
	\end{figure}

In the frequency clustering process, NMD algorithm fails to decompose any modes when the energy threshold $P=90\%$. We gradually increase the energy proportion $P$ to $98\%$, and obtain different decomposition results, shown in Fig. \ref{Hu6} and Fig. \ref{Hu7}. Due to the big energy difference between components of $x_{P}(t)$, it needs a larger $P$ to separate high frequency from low frequency. As $P$ increasing, frequency 2, frequency 24, frequency 100 are selected as primary frequency component and separated as a single mode in turn. As for the noise, we can find in Fig. \ref{Hu6} and Fig. \ref{Hu7} that the noise evenly appears in every mode, and the energy of noise uniformly distribute in every frequency.

\begin{figure}
		\begin{subfigure}[t]{0.23\textwidth}
			\centering
			\includegraphics[width=\textwidth]{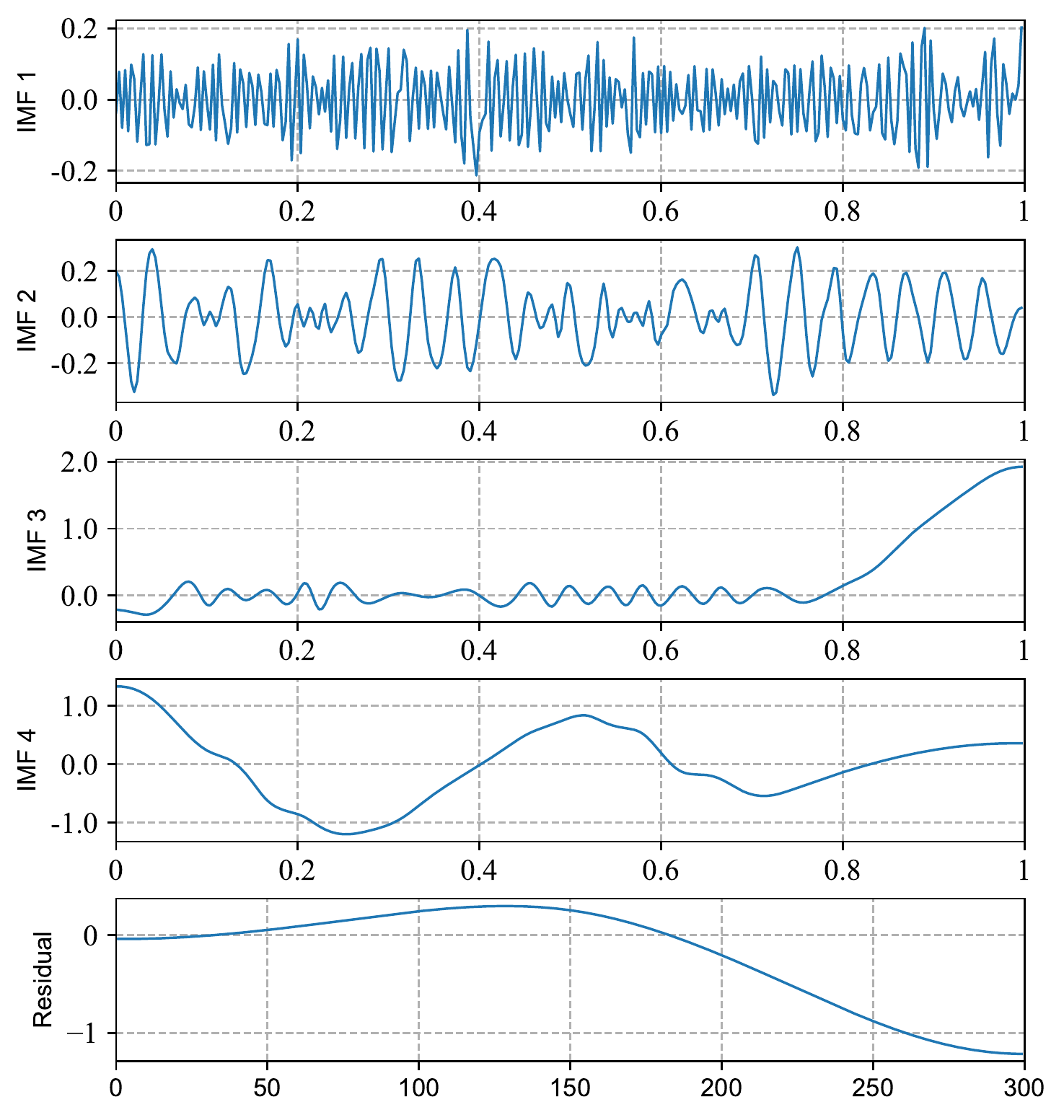}
            \label{Hu8a}
            \vspace{-2em}
			\caption{EMD }
		\end{subfigure}
		\begin{subfigure}[t]{.23\textwidth}
			\centering
			\includegraphics[width=\textwidth]{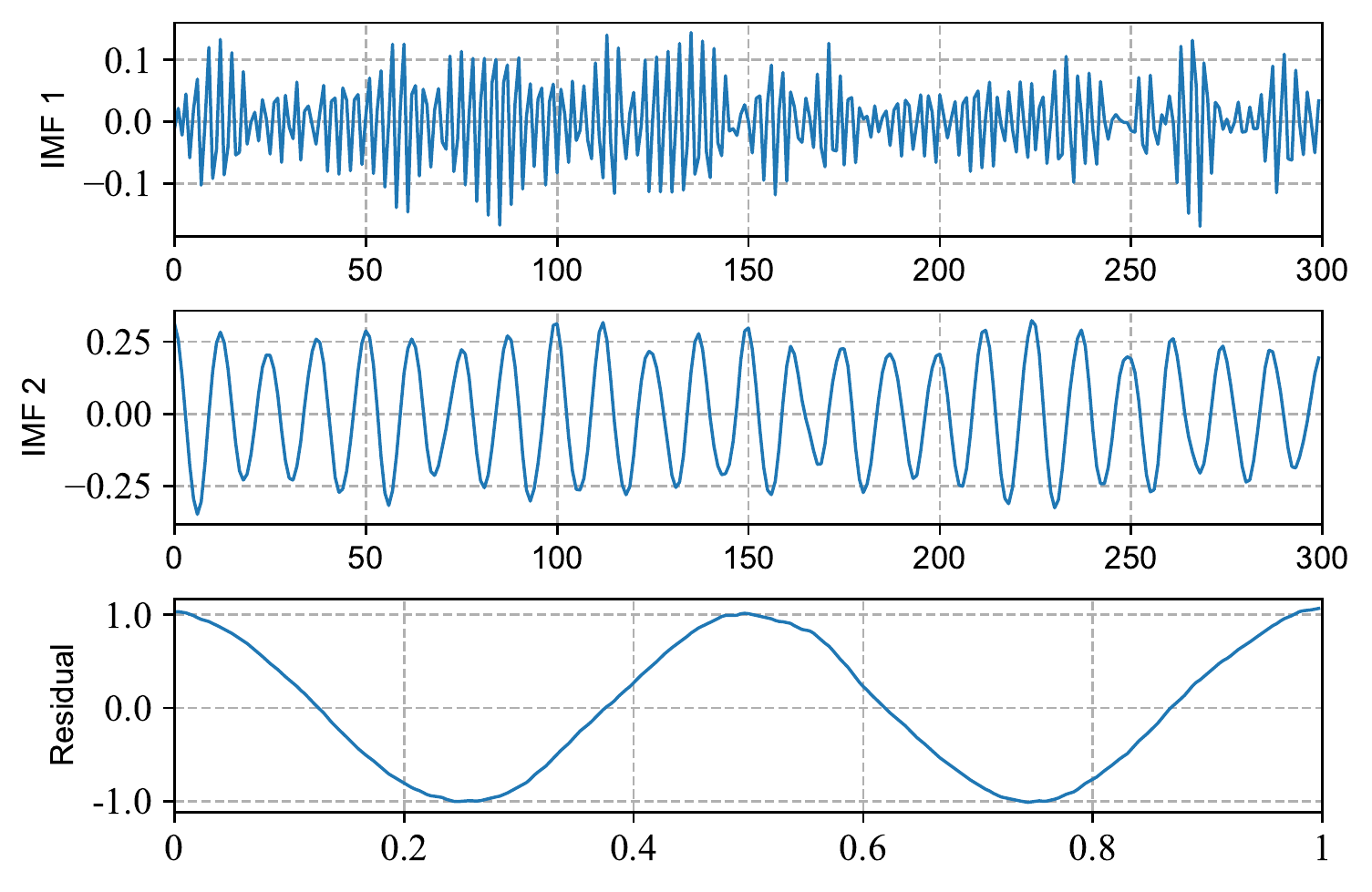}
            \label{Hu8b}
            \vspace{-2em}
			\caption{VMD}
		\end{subfigure}
        \vspace{-1em}
        \caption{Decompositions of EMD and VMD on $x_{P}(t)$. (a) EMD's results. (b) VMD's results.}
        \vspace{-1em}
        \label{Hu8}
	\end{figure}

We also utilize EMD and VMD to decompose $x_{P}(t)$ and illustrate the results in Fig. \ref{Hu8}. EMD results distort the original frequency components, and hardly separate any complete component because of the interference of noise. In VMD results, frequency 8 and 2 are completely recovered, but frequency 24 presents a totally wrong amplitude. In fact, VMD tends to divide the noise into residual on account of the compact supported frequency domain of its IMFs. In summary, NMD can recover original components under noise but perform worse data denoising than VMD, and EMD can hardly recover any complete original components under strong noise.

\subsection{Comparison and Analysis of decomposition results of EMD, VMD and NMD}\label{sec3.2}

For further studying the validity of our approach, we choose 4 groups of datas to compare NMD with EMD and VMD, which are most widely used in mode decomposition. Among 4 datas, two synthetic datas used in \cite{dragomiretskiy2013variational} with muti-components and special frequency modulation and amplitude modulation are selected to intuitively compare the decomposition results; besides, two real datasets, monthly totals of national airline passengers in American \cite{Airline} and monthly ozone concentration in downtown Los Angeles \cite{godfrey2017neural} are also selected to reflect the practical application of decomposition algorithms. We first present the detail decomposition results and make a overall comparison, in which the ability of collecting features, end effect, spectral mixing, et. al are mainly considered. Orthogonality and completeness are important criteria being considered in all kinds of decomposition algorithms. Thus, we then specially list the the overall index of orthogonality (IO) and the mean absolute error (MAE) between original data and sum of decomposed components, and analyze the orthogonality and reconstruction error in isolation.

Specially, considering the ability of extracting information on 2 real datas, we calculate the proportion of mode energy on original data Percentage Energy (PE) and the correlation coefficient between components and original data to evaluate, in which the PE of the $k$th mode is expressed as
\begin{equation}\label{eq6}
PE=\frac{\sum^{N-1}_{t=0}u^{2}_{i}(t)}{\sum^{K+1}_{k=1}\sum^{N-1}_{t=0}u^{2}_{i}(t)},i=1,2,...K+1,
\end{equation}
where $u_{k}(t)$ is the $k$th IMF of decomposition components and $u_{K+1}(t)$ is the residual.

\subsubsection{Decomposition Results}\label{sec3.2.1}
{}

\begin{figure}
		\begin{subfigure}[t]{0.15\textwidth}
			\centering
			\includegraphics[width=\textwidth]{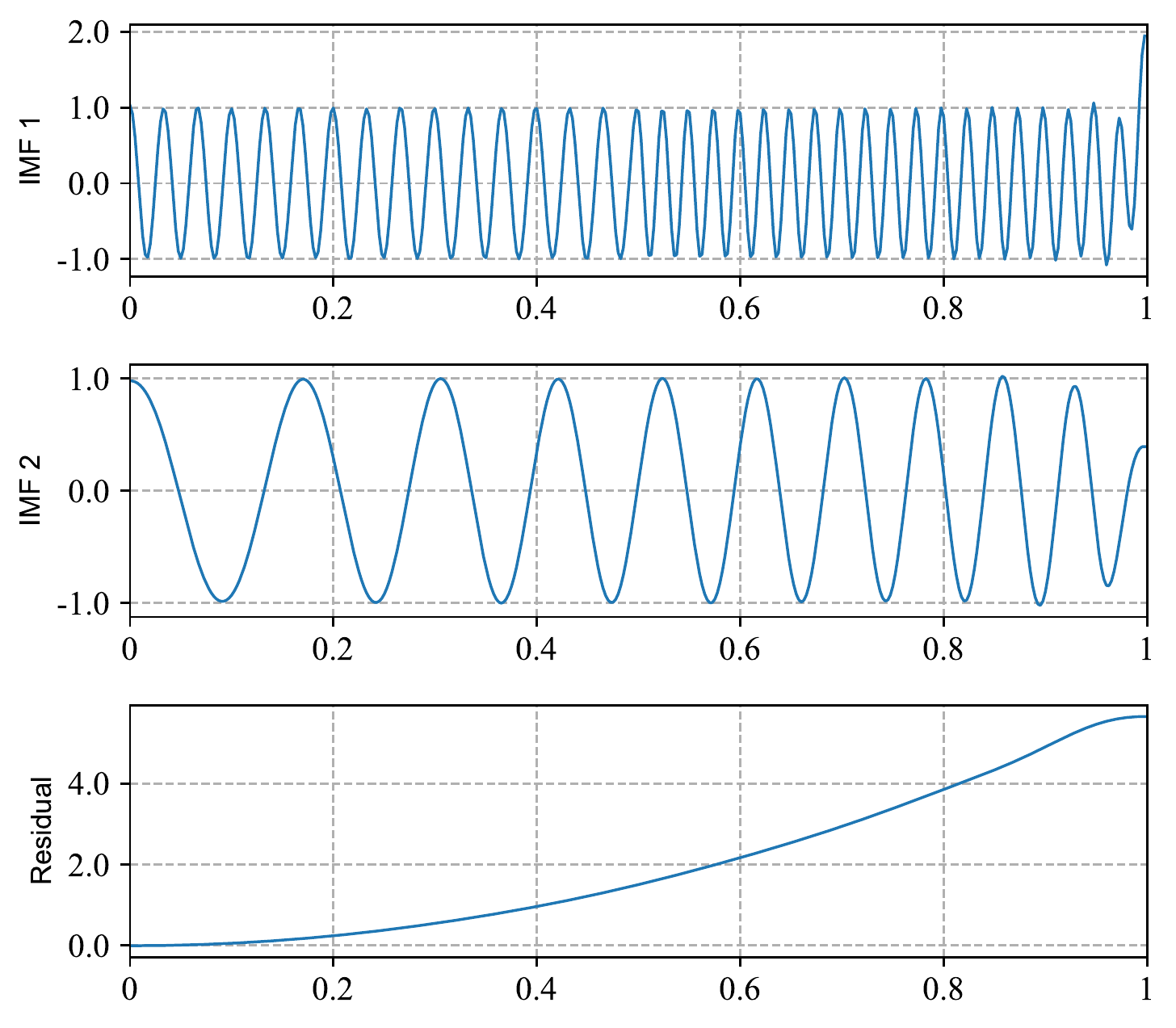}
            \label{Hu9a}
            \vspace{-2em}
			\caption{EMD}
		\end{subfigure}
		\begin{subfigure}[t]{.15\textwidth}
			\centering
			\includegraphics[width=\textwidth]{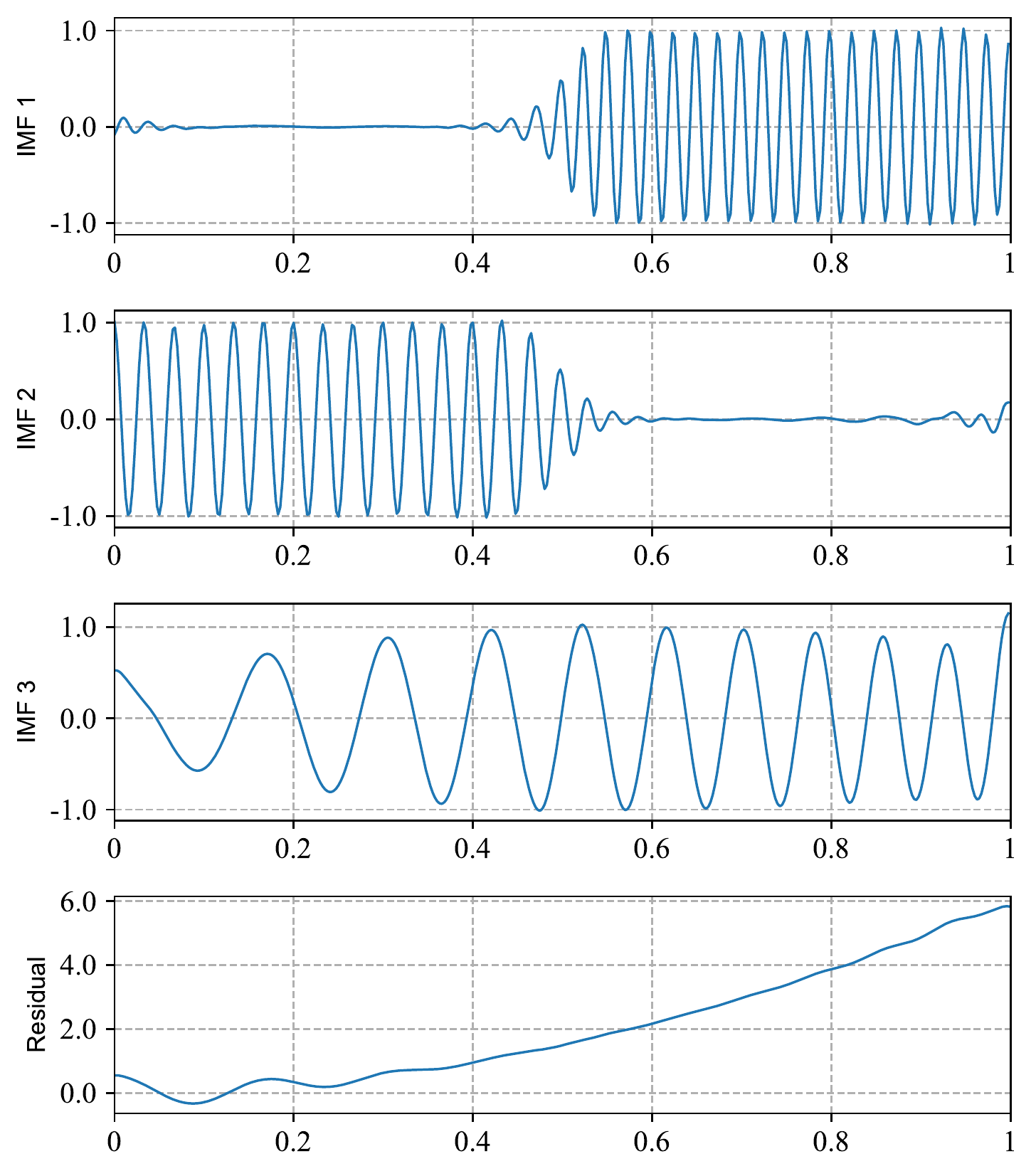}
            \label{Hu9b}
            \vspace{-2em}
			\caption{VMD}
		\end{subfigure}
		\begin{subfigure}[t]{.15\textwidth}
			\centering
			\includegraphics[width=\textwidth]{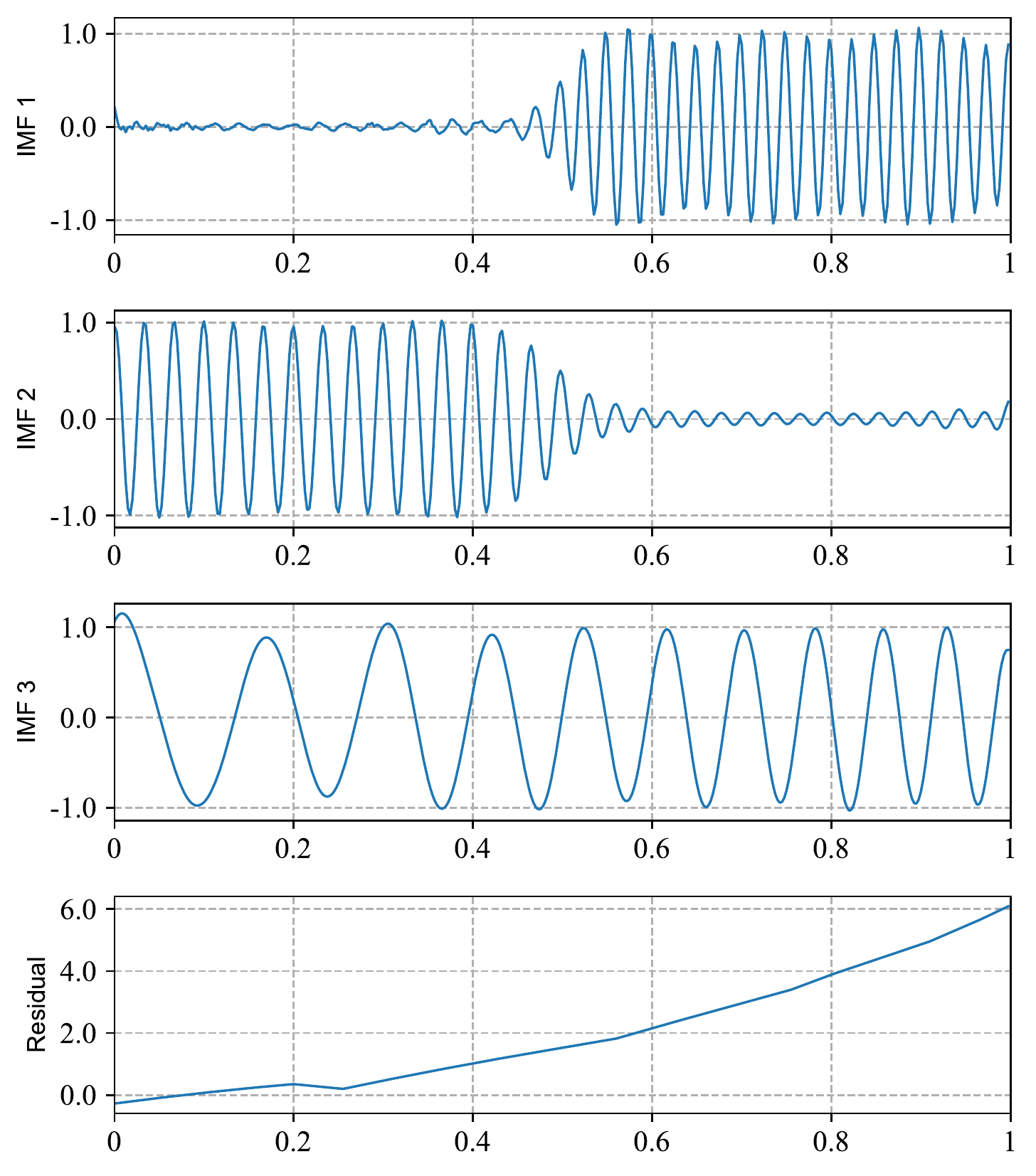}
            \label{Hu9c}
            \vspace{-2em}
			\caption{NMD}
		\end{subfigure}
        \vspace{-1em}
        \caption{Decompositions of EMD, VMD and NMD on $x_{1}(t)$. (a) EMD's results. (b) VMD's results. (c) NMD's results.}
        \vspace{-1em}
        \label{Hu9}
	\end{figure}

\emph{1) $x_{1}(t)$}: The first data including a quadratic trend, a chirp signal and a component with sharp transition between two constant frequencies. Fig. \ref{Hu9} illustrate the decomposition results of EMD, VMD and NMD on this data.
\begin{equation}\label{eq7}
x_{1}(t)=6t^{2}+\cos(10\pi t+10\pi t^{2})+\left\{
             \begin{aligned}
             &\cos(60\pi t), t\leq0.5   \\
             &\cos(80\pi t-10\pi),t>0.5
             \end{aligned}
             \right.
\end{equation}

Fig. \ref{Hu9} shows that the non-periodic component can be roughly separated in three decompositions, but all appear different degree of distortion at the endpoints, in which the EMD distorts the trend at the right end of interval, NMD distorts the trend at the left endpoints, and VMD distorts both endpoints. This end effect is further reflected in IMF of $\cos(10\pi t+10\pi t^{2})$, because some part of this component is mixed into the residual at the endpoints, among them VMD's end effect is the most serious. Two linear frequency modulation part with finite time domin in the third component of $x_{1}(t)$ manage to be separated into two modes in VMD and NMD results, where the separation of VMD is more complete than NMD. EMD fails to separate two noncontinuous frequency of the third component in its results.

\begin{figure}
		\begin{subfigure}[t]{0.15\textwidth}
			\centering
			\includegraphics[width=\textwidth]{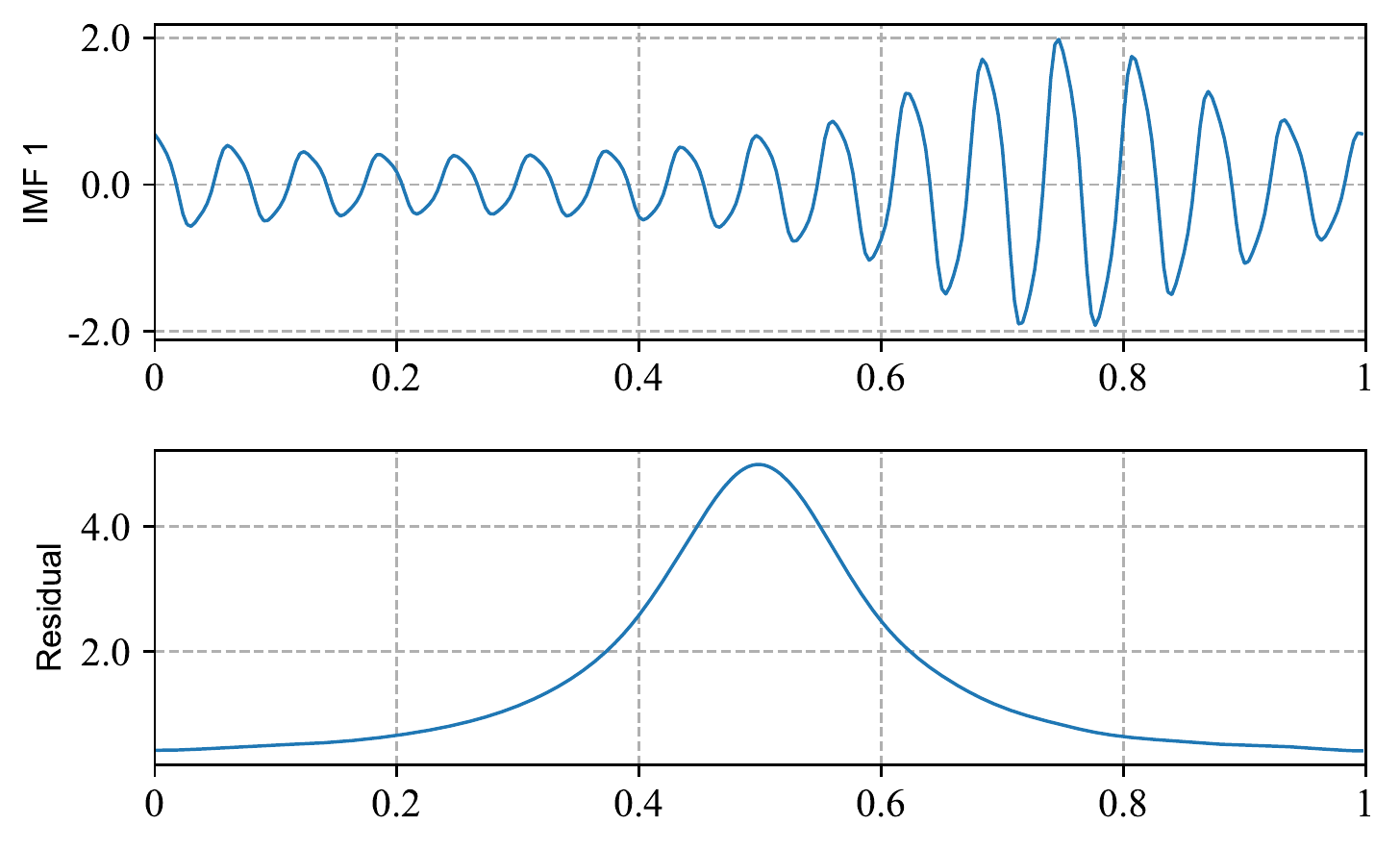}
            \label{Hu10a}
            \vspace{-2em}
			\caption{EMD}
		\end{subfigure}
		\begin{subfigure}[t]{.15\textwidth}
			\centering
			\includegraphics[width=\textwidth]{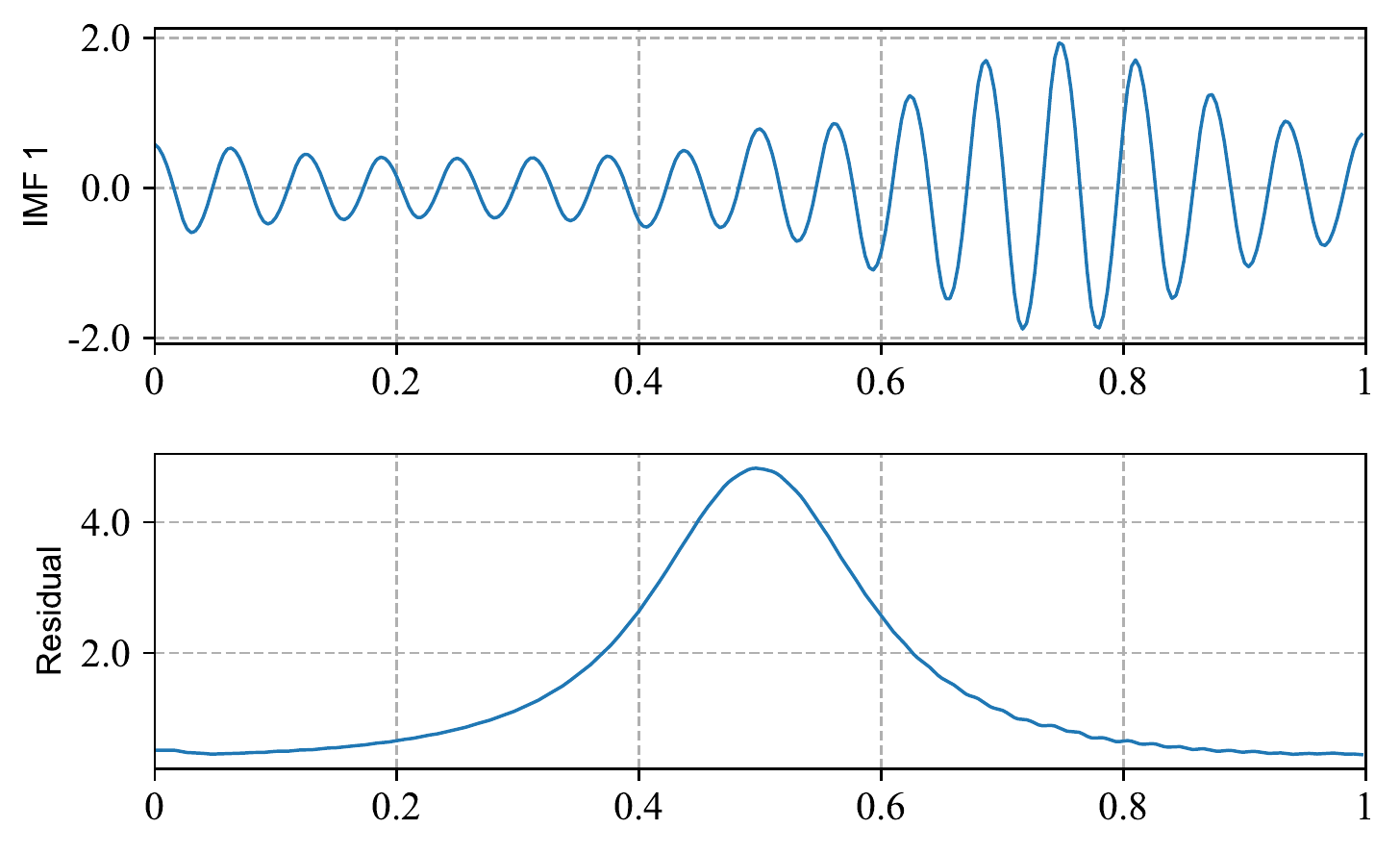}
            \label{Hu10b}
            \vspace{-2em}
			\caption{VMD}
		\end{subfigure}
		\begin{subfigure}[t]{.15\textwidth}
			\centering
			\includegraphics[width=\textwidth]{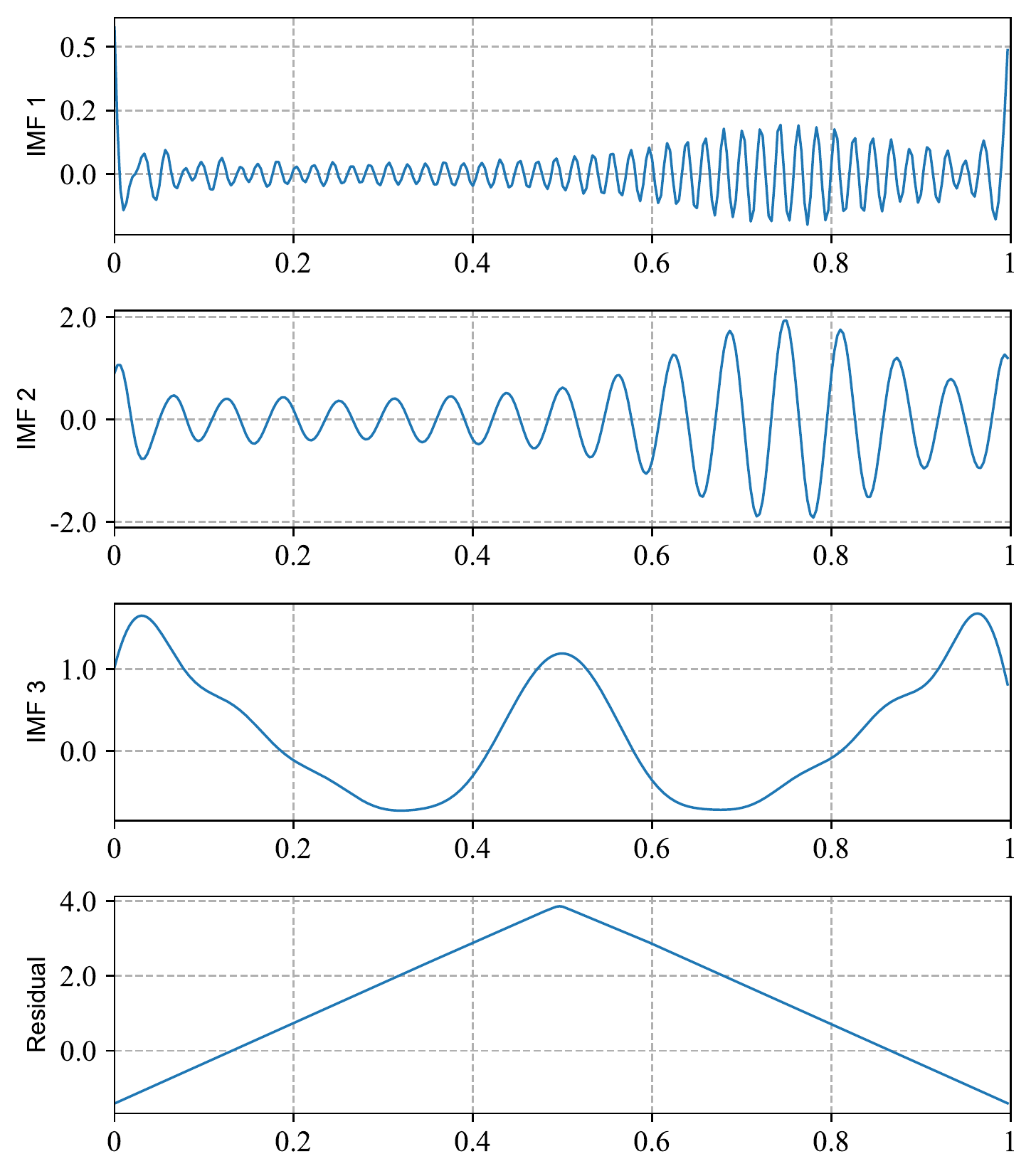}
            \label{Hu10c}
            \vspace{-2em}
			\caption{NMD}
		\end{subfigure}
        \vspace{-1em}
        \caption{Decompositions of EMD, VMD and NMD on $x_{2}(t)$. (a) EMD's results. (b) VMD's results. (c) NMD's results.}
        \label{Hu10}
        \vspace{-1em}
	\end{figure}

\emph{2) $x_{2}(t)$}: The second data composed of a bell-shaped low frequency component and a high frequency AM-FM component. The decompositions of EMD, VMD and NMD on this data are shown in Fig. \ref{Hu10}.
\begin{equation}\label{eq7}
x_{2}(t)=\frac{1}{1.2+\cos(2\pi t)}+\frac{\cos(32\pi t+0.2\cos(64\pi t))}{1.5+\sin(2\pi t)}
\end{equation}

In Fig. \ref{Hu10}, EMD and VMD present a similar processing with two decomposed modes, while NMD obtain 4 modes. The first low frequency component of $x_{2}(t)$ is actually periodic, but whose period only occurs once in selected time domain $[0,1]$. In EMD and VMD, this component fails to be recognized as periodic component but as residual component. NMD distinguish that there has periodicity in the low frequency component, but do not recover its real periodicity due to the insufficient sampling, further resulting a wrong illustration on residual and the endpoints of IMF1 and IMF2. NMD get another IMF corresponding to the frequency modulation part of the high frequency component of $x_{2}(t)$, which is not recovered by EMD and VMD. These results indicate that NMD has a better ability on feature mining than EMD and VMD.

\begin{figure}
  \centering
  \includegraphics[width=2in]{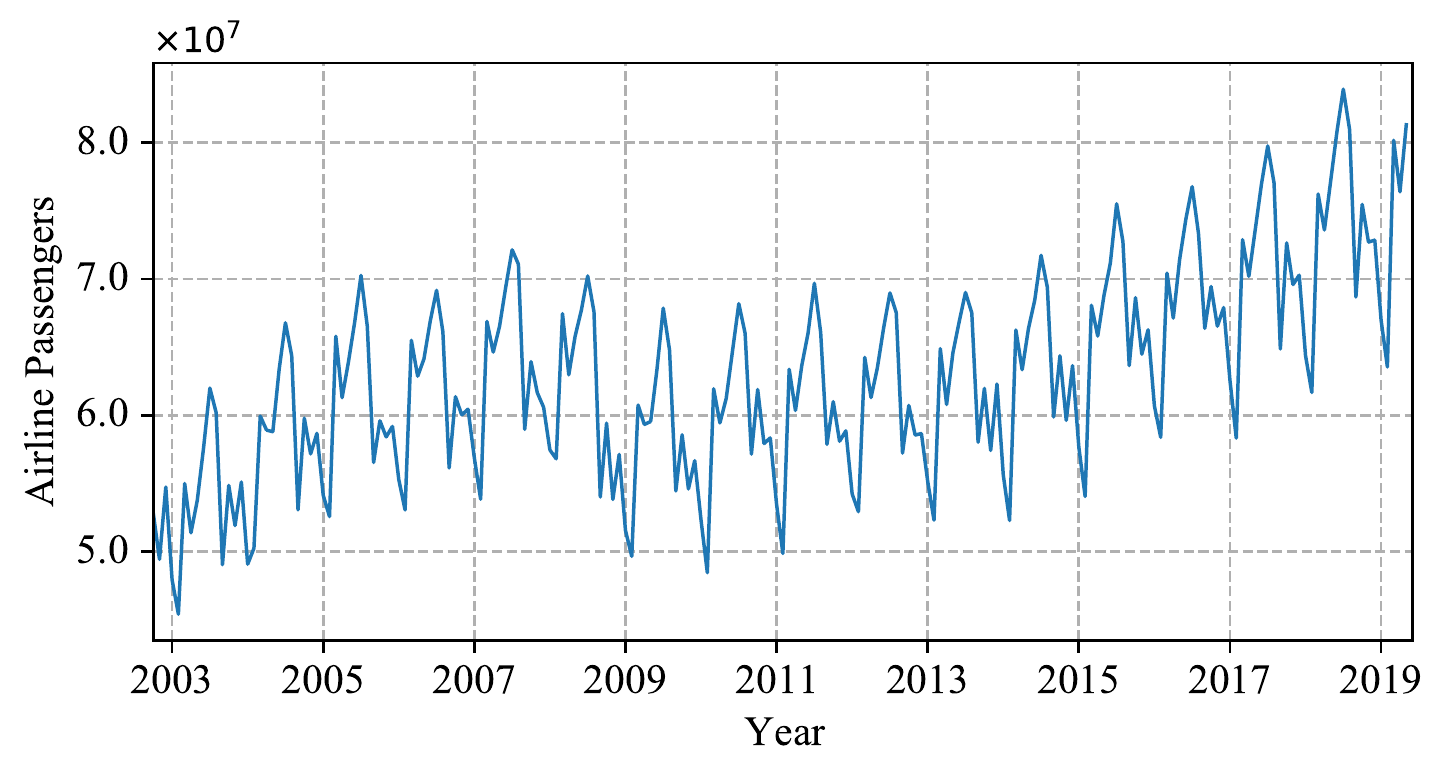}
  \vspace{-1.5em}
  \caption{Monthly totals of national airline passengers in American from October 2002 to August 2018.\label{Hu11}}
  \vspace{-1em}
\end{figure}

\emph{3) Airline Passengers data}: This data is the monthly totals of national airline passengers in American from October 2002 to August 2018, illustrated in Fig. \ref{Hu11}. The passenger flow data usually has typical periodicity, i.e., year, quarter and month.

\begin{figure}
		\begin{subfigure}[t]{0.155\textwidth}
			\centering
			\includegraphics[width=\textwidth]{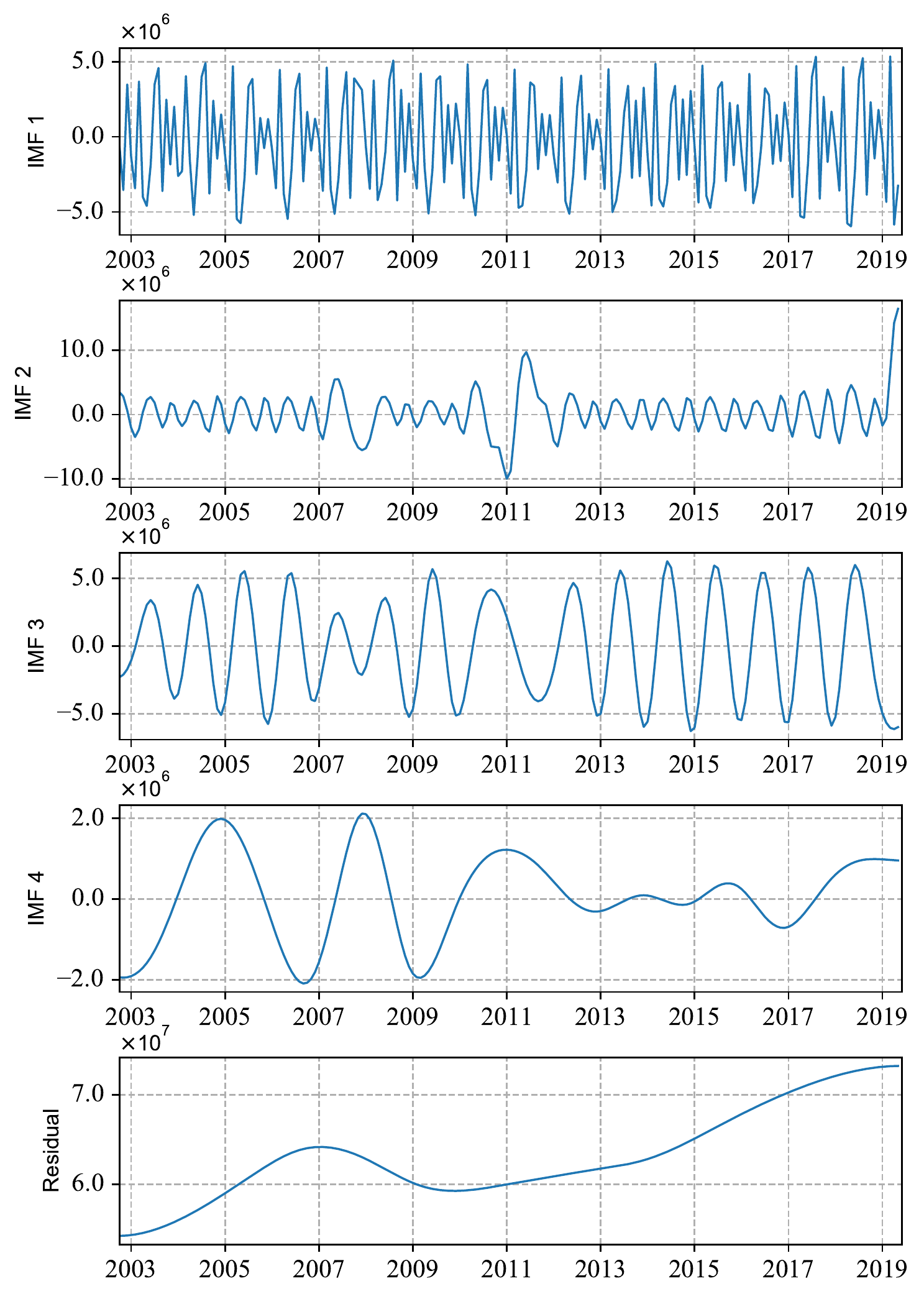}
            \label{Hu12a}
            \vspace{-2em}
			\caption{EMD}
		\end{subfigure}
		\begin{subfigure}[t]{.155\textwidth}
			\centering
			\includegraphics[width=\textwidth]{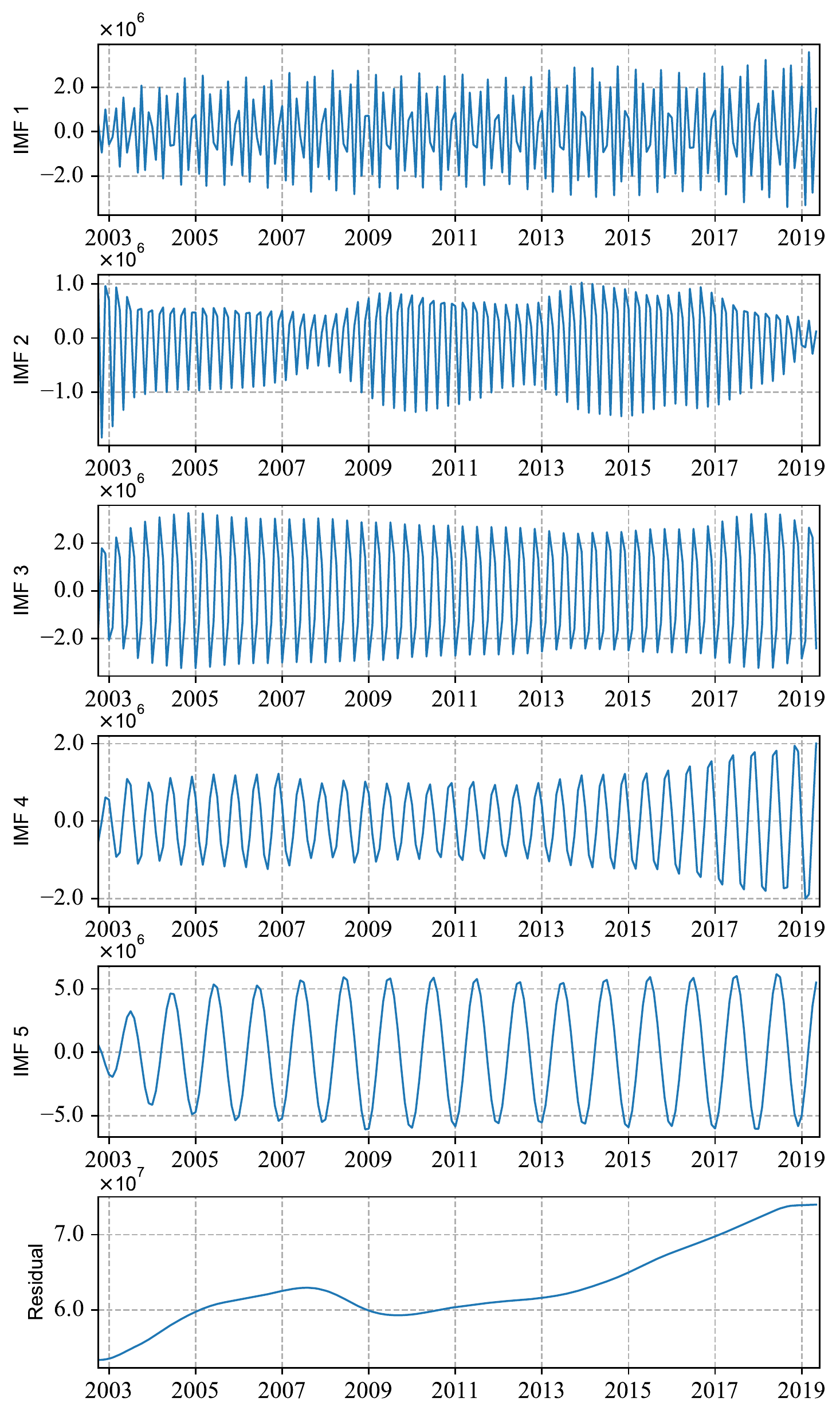}
            \label{Hu12b}
            \vspace{-2em}
			\caption{VMD}
		\end{subfigure}
		\begin{subfigure}[t]{.155\textwidth}
			\centering
			\includegraphics[width=\textwidth]{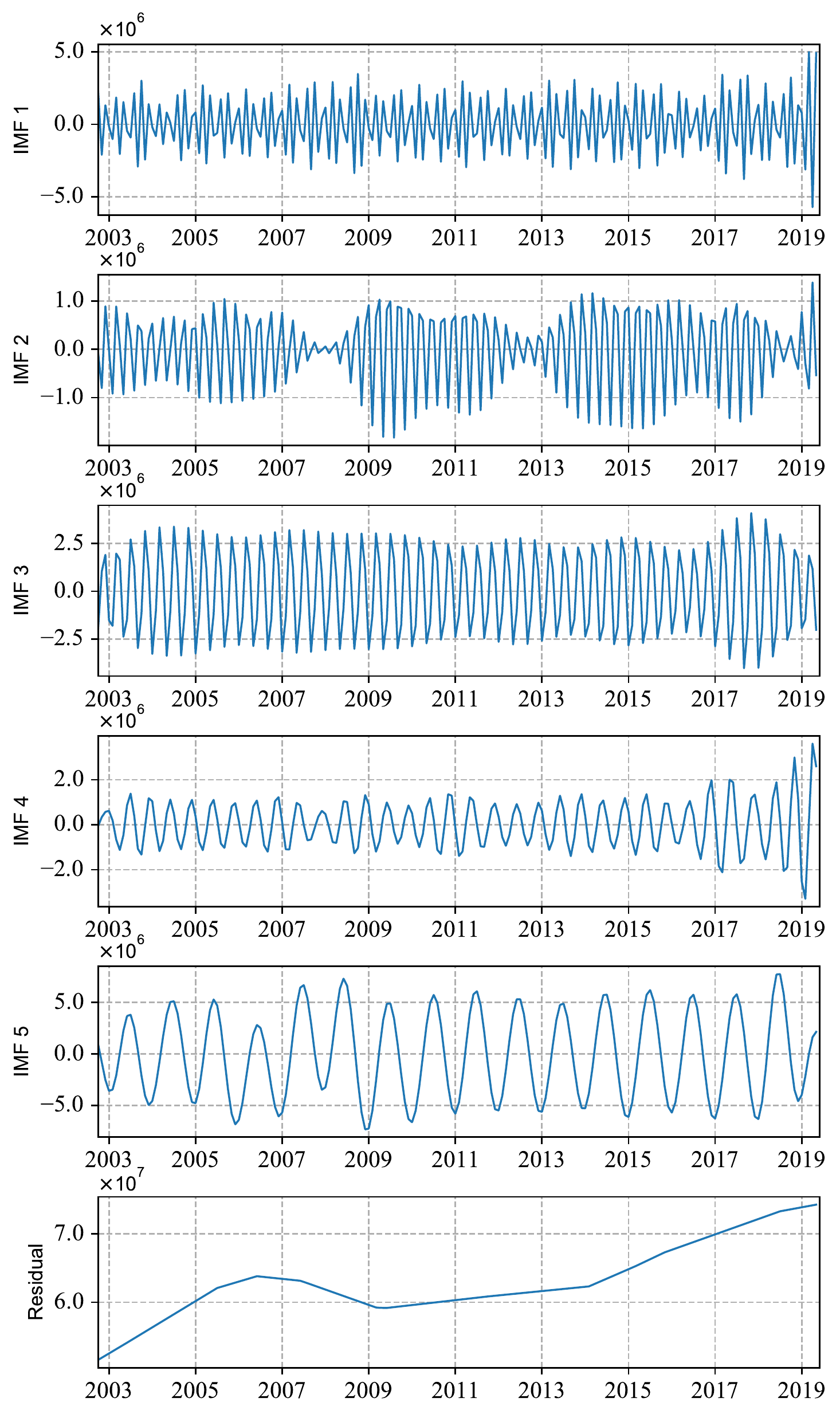}
            \label{Hu12c}
            \vspace{-2em}
			\caption{NMD}
		\end{subfigure}
        \vspace{-1em}
        \caption{Decompositions of EMD, VMD and NMD on Airline data. (a) EMD's results. (b) VMD's results. (c) NMD's results.}
        \vspace{-1em}
        \label{Hu12}
	\end{figure}

The decompositions of EMD, VMD and NMD are shown in Fig. \ref{Hu12}. EMD, VMD and NMD divides the original data into 5, 6 and 6 modes respectively, and the corresponding periods of IMFs are about 2, 3, 4, 6 and 12 months. The passenger flow of national airlines shows a growth trend over the long term, but decrease a lot over 2005 to 2007. It can be inferred that some special events occurred during this period, causing the abnormal trend in a short time. The IMFs of EMD results are all mixed with other mode components. The VMD and NMD results are close, all completely recovering several main periodic components of the Airline data, only kind of different on their amplitudes. The analysis of VMD and NMD needs to be further investigated by PE and correlation coefficient, as listed in Table. \ref{T1}.
The PE of IMFs of EMD are uniform with IMF1-3, while that of IMF1-4 in EMD and NMD are much smaller than IMF5, because the related frequency component of IMF5 is shared by EMD's IMF1-3. The PE and correlation coefficient of long term trend in three methods are essentially the same, suggesting that the long term trend is correctly recovered in all three decomposition results. IMF3 and IMF5's PE and correlation coefficient in VME and NMD results all are the two largest, indicating that the period of these two modes, 4 and 12 months are two most meaningful components of Airline data.
\begin{table}
\caption{PE and correlation coefficient of modes and original data on Airline data.}
\vspace{-1em}
\label{T1}
\begin{minipage}{\columnwidth}
\begin{center}
\begin{tabular}{cccccccc}
\hline
\multicolumn{2}{c}{}& IMF1&IMF2&IMF3&IMF4&IMF5&Res\\\hline
 \multirow{2}*{EMD} &PE&0.28&0.26&0.33&0.03&/&99.10\\
\cline{2-8}
~&Corr&0.32&0.38&0.40&0.23&/&0.70\\\hline
 \multirow{2}*{VMD}&PE&0.09&0.01&0.12&0.02&0.39&99.35\\
\cline{2-8}
~&Corr&0.26&0.12&0.30&0.14&0.55&0.71\\\hline
 \multirow{2}*{NMD}&PE&0.10&0.01&0.19&0.02&0.40&99.34\\
\cline{2-8}
~&Corr&0.27&0.10&0.29&0.16&0.55&0.71\\\hline
\end{tabular}
\end{center}
\vspace{-1em}
\end{minipage}
\end{table}

\begin{figure}
		\begin{subfigure}[t]{0.2\textwidth}
			\centering
			\includegraphics[width=\textwidth]{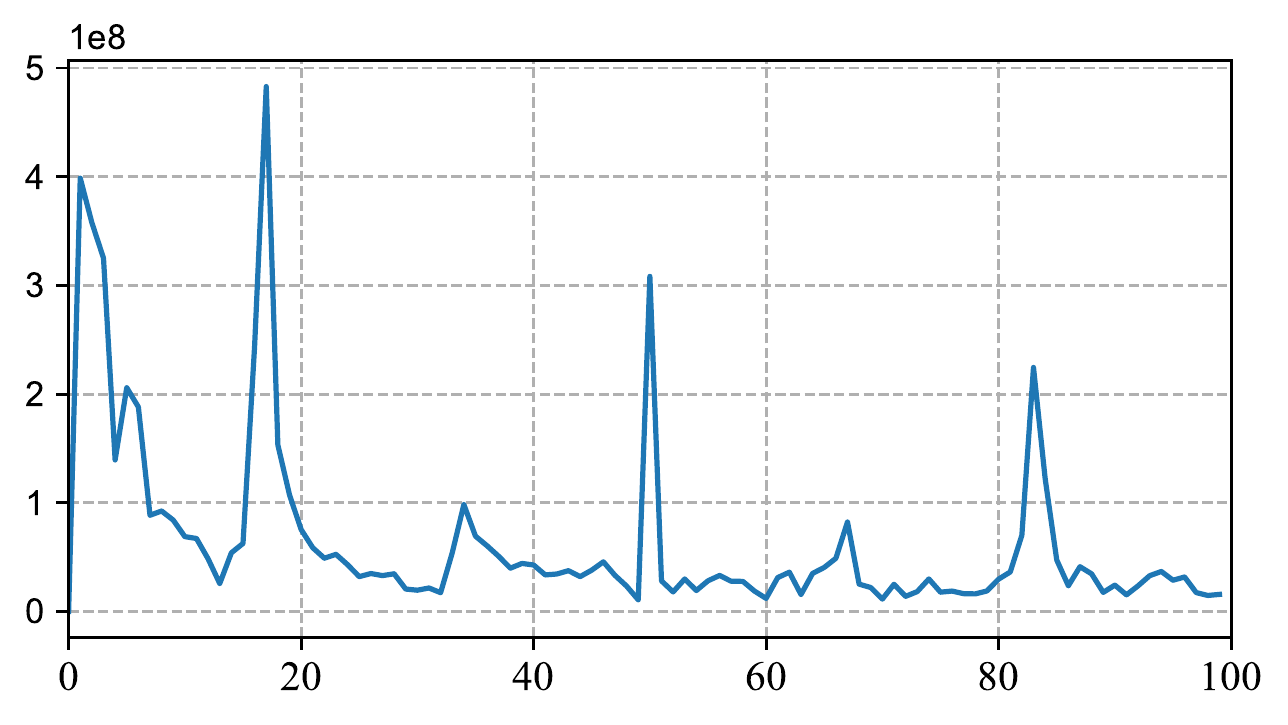}
            \label{Hu13a}
            \vspace{-2em}
			\caption{Airline data}
		\end{subfigure}
		\begin{subfigure}[t]{.2\textwidth}
			\centering
			\includegraphics[width=\textwidth]{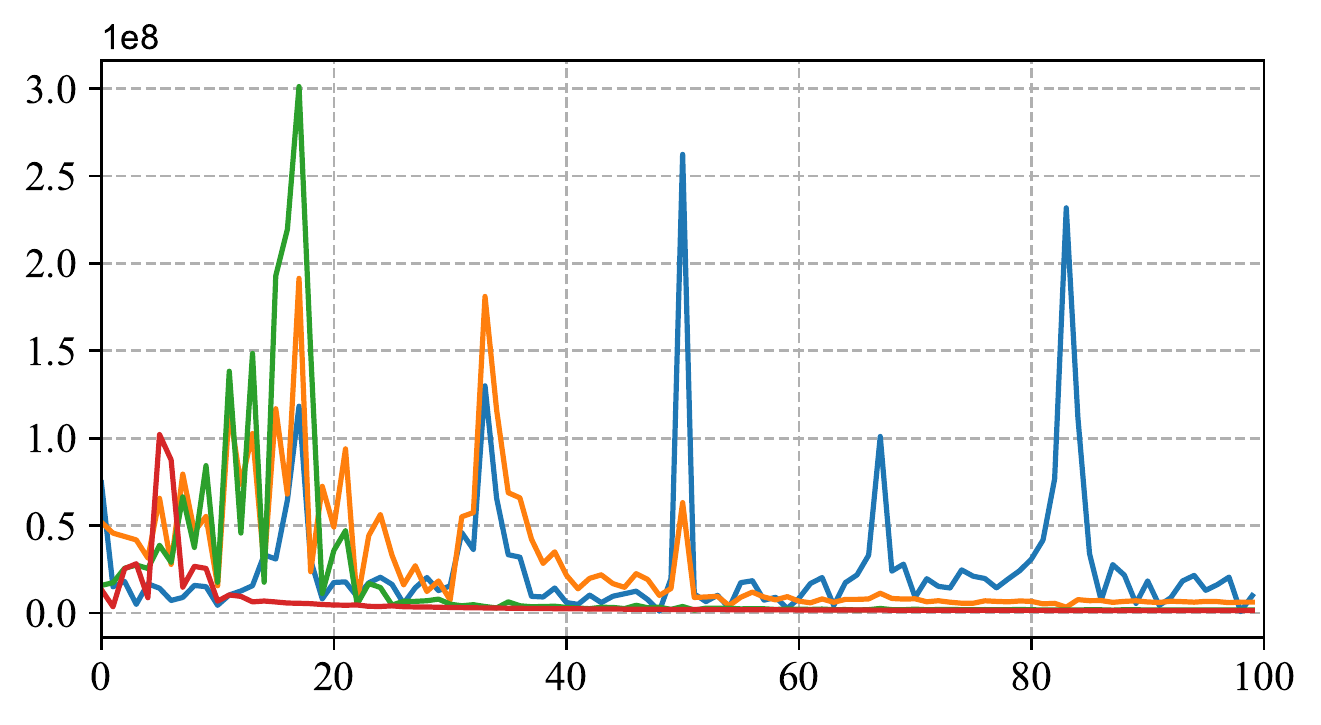}
            \label{Hu13b}
            \vspace{-2em}
			\caption{EMD}
		\end{subfigure}
		\begin{subfigure}[t]{.2\textwidth}
			\centering
			\includegraphics[width=\textwidth]{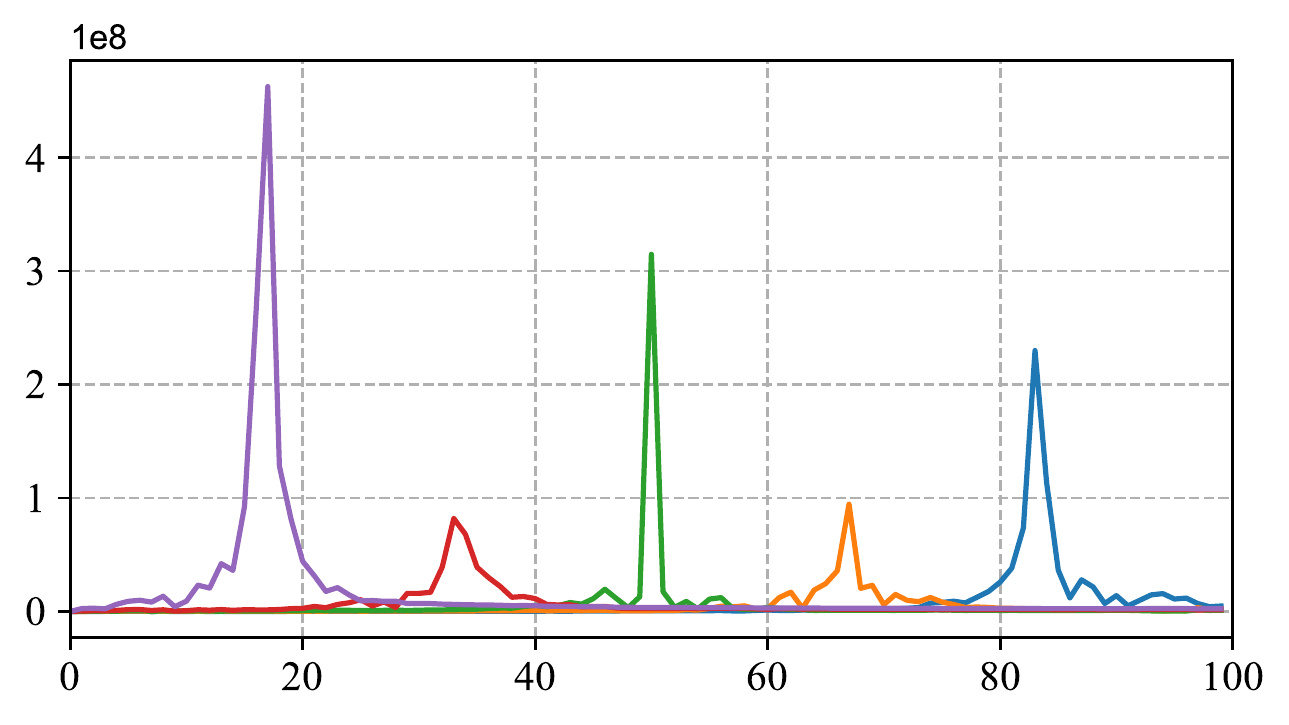}
            \label{Hu13c}
            \vspace{-2em}
			\caption{VMD}
		\end{subfigure}
		\begin{subfigure}[t]{.2\textwidth}
			\centering
			\includegraphics[width=\textwidth]{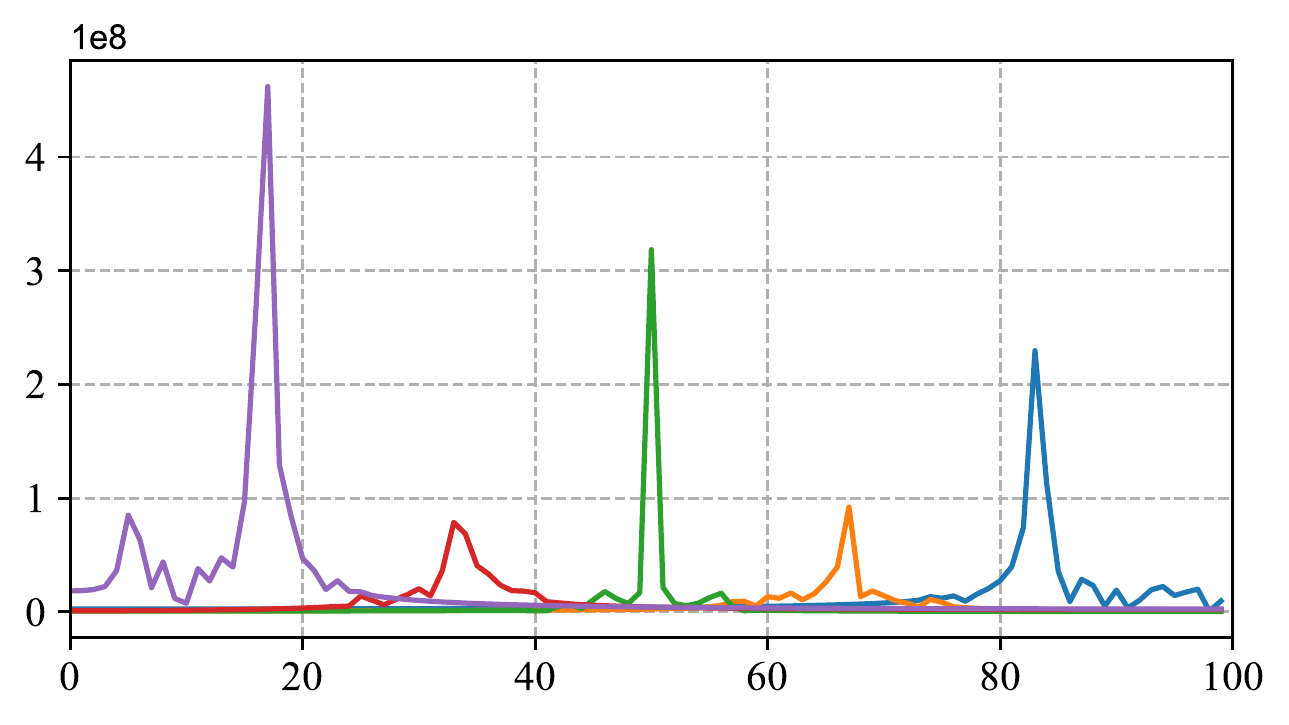}
            \label{Hu13d}
            \vspace{-2em}
			\caption{NMD}
		\end{subfigure}
        \vspace{-1em}
        \caption{Spectrum of Airline data and its decompositions. (a) Spectrum of zero-mean normalized Airline data. (b) Spectrum of EMD results. (c) Spectrum of VMD results. (d) Spectrum of NMD results.}
        \vspace{-1em}
        \label{Hu13}
	\end{figure}

We further apply DFT on original data and the IMFs decomposed by the algorithms to obtain its spectrum, as shown in Fig. \ref{Hu13}. The DFT on original data is carried with which subtracts its average. In EMD results, there are a serious frequency mixing between each mode. The spectrum of VMD components are all compactly supported near the central frequency, indicating that VMD's IMFs may lose some small frequency events far from central frequency on spectrum. NMD tends to fit all frequency components of data and evenly divide the random components into each IMF.

\begin{figure}
  \centering
  \includegraphics[width=2in]{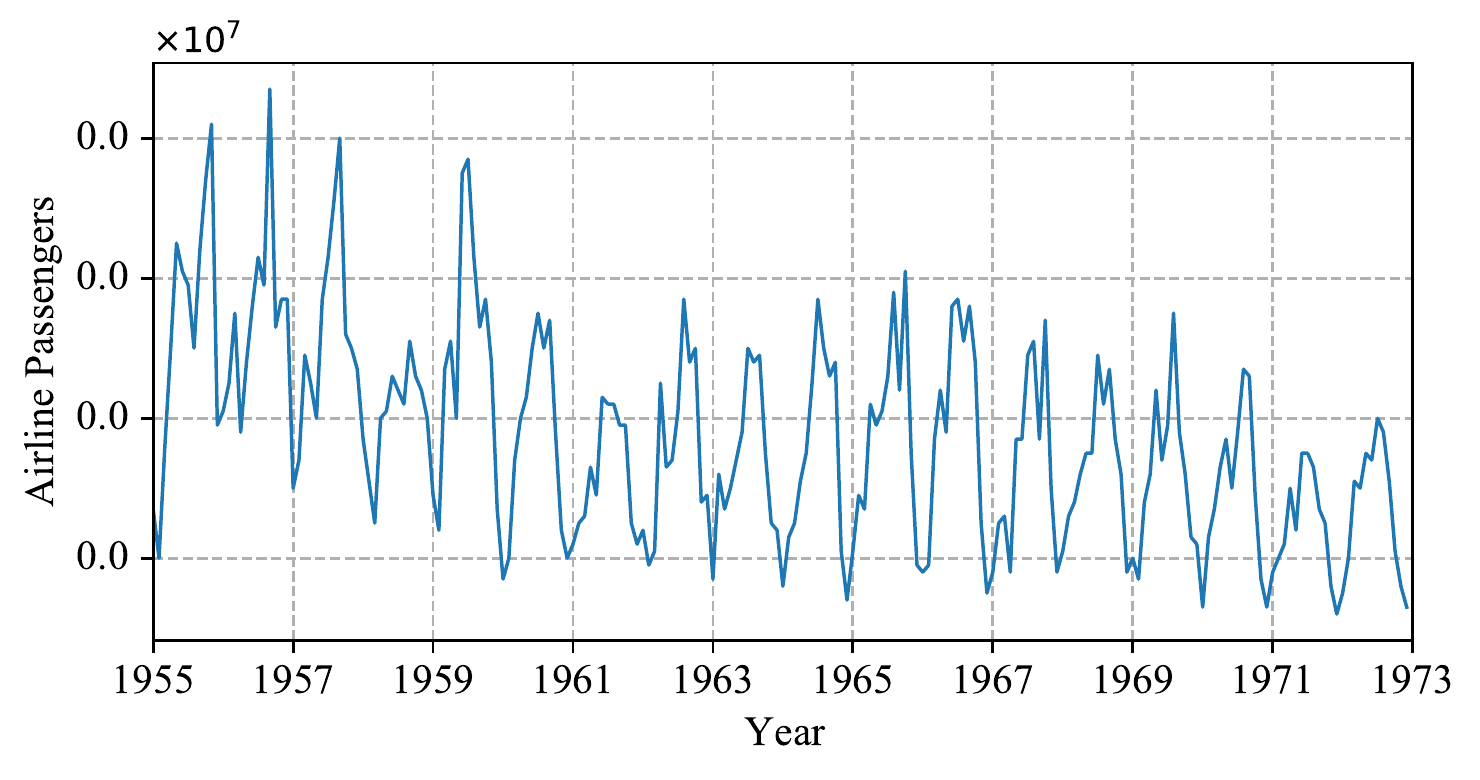}
  \vspace{-1em}
  \caption{Monthly ozone concentration in downtown Los Angeles from January 1955 to August 1967.\label{Hu14}}
\end{figure}

\begin{figure}
		\begin{subfigure}[t]{0.15\textwidth}
			\centering
			\includegraphics[width=\textwidth]{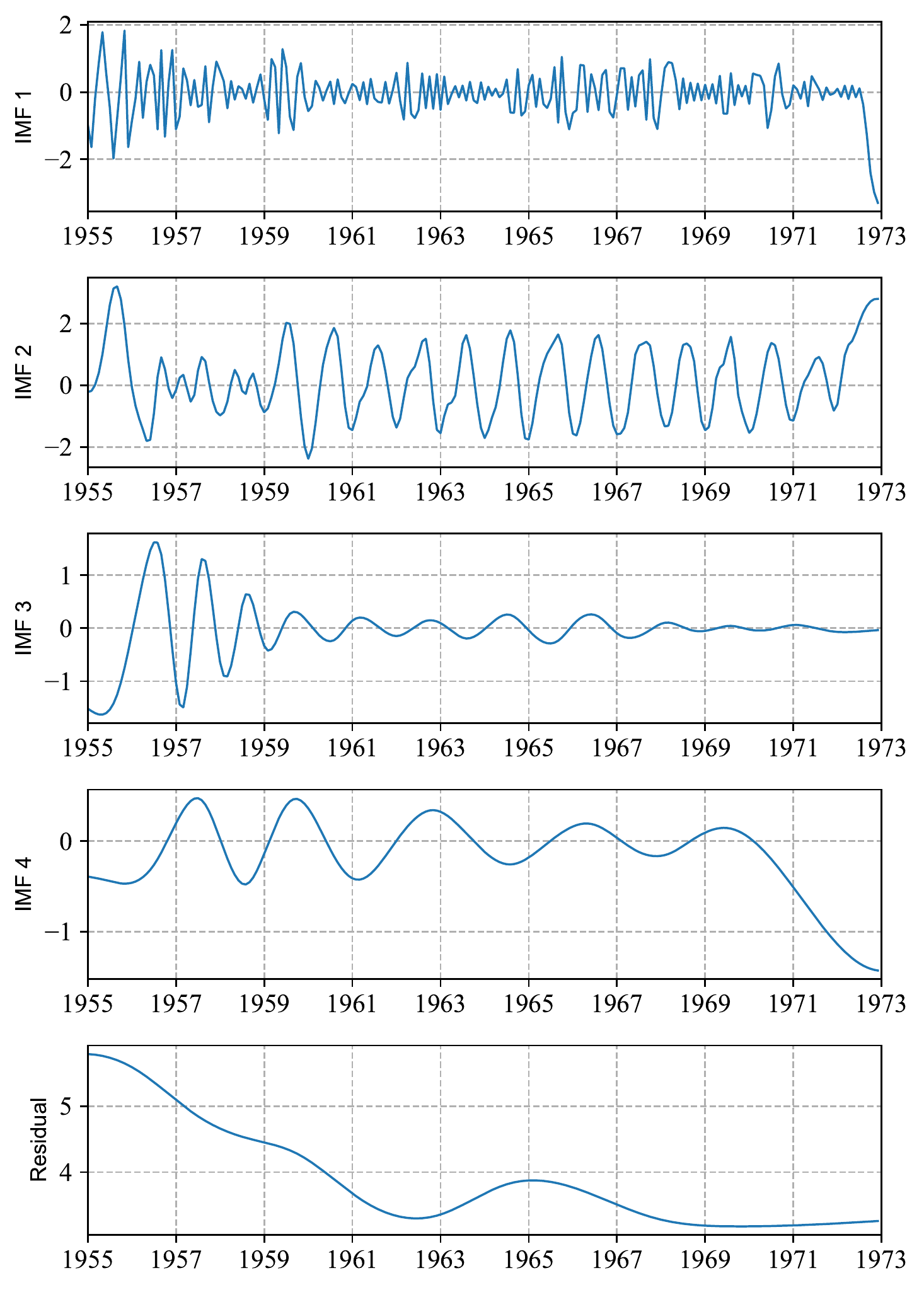}
            \label{Hu15a}
            \vspace{-2em}
			\caption{EMD}
		\end{subfigure}
		\begin{subfigure}[t]{.15\textwidth}
			\centering
			\includegraphics[width=\textwidth]{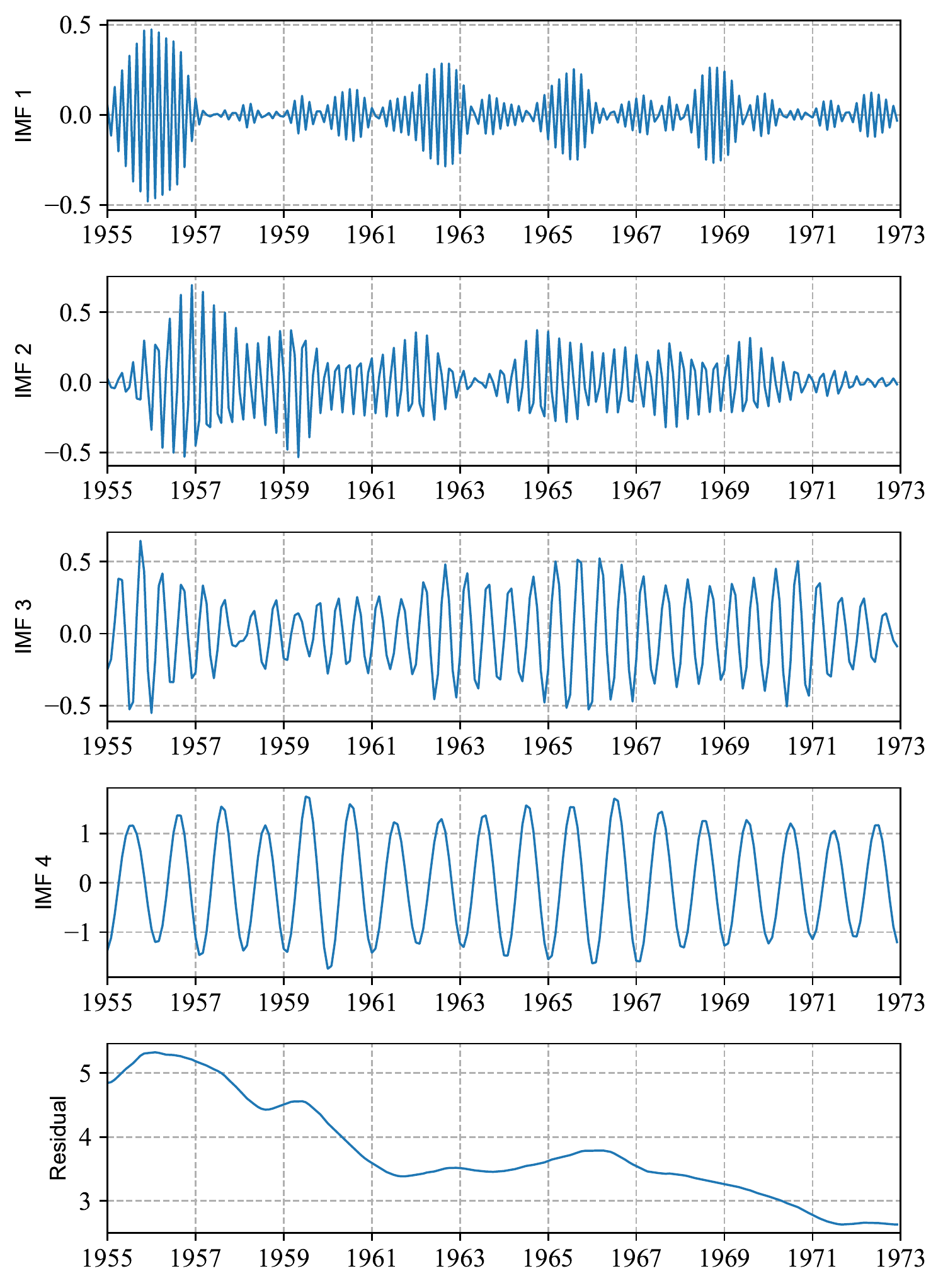}
            \label{Hu15b}
            \vspace{-2em}
			\caption{VMD}
		\end{subfigure}
		\begin{subfigure}[t]{.15\textwidth}
			\centering
			\includegraphics[width=\textwidth]{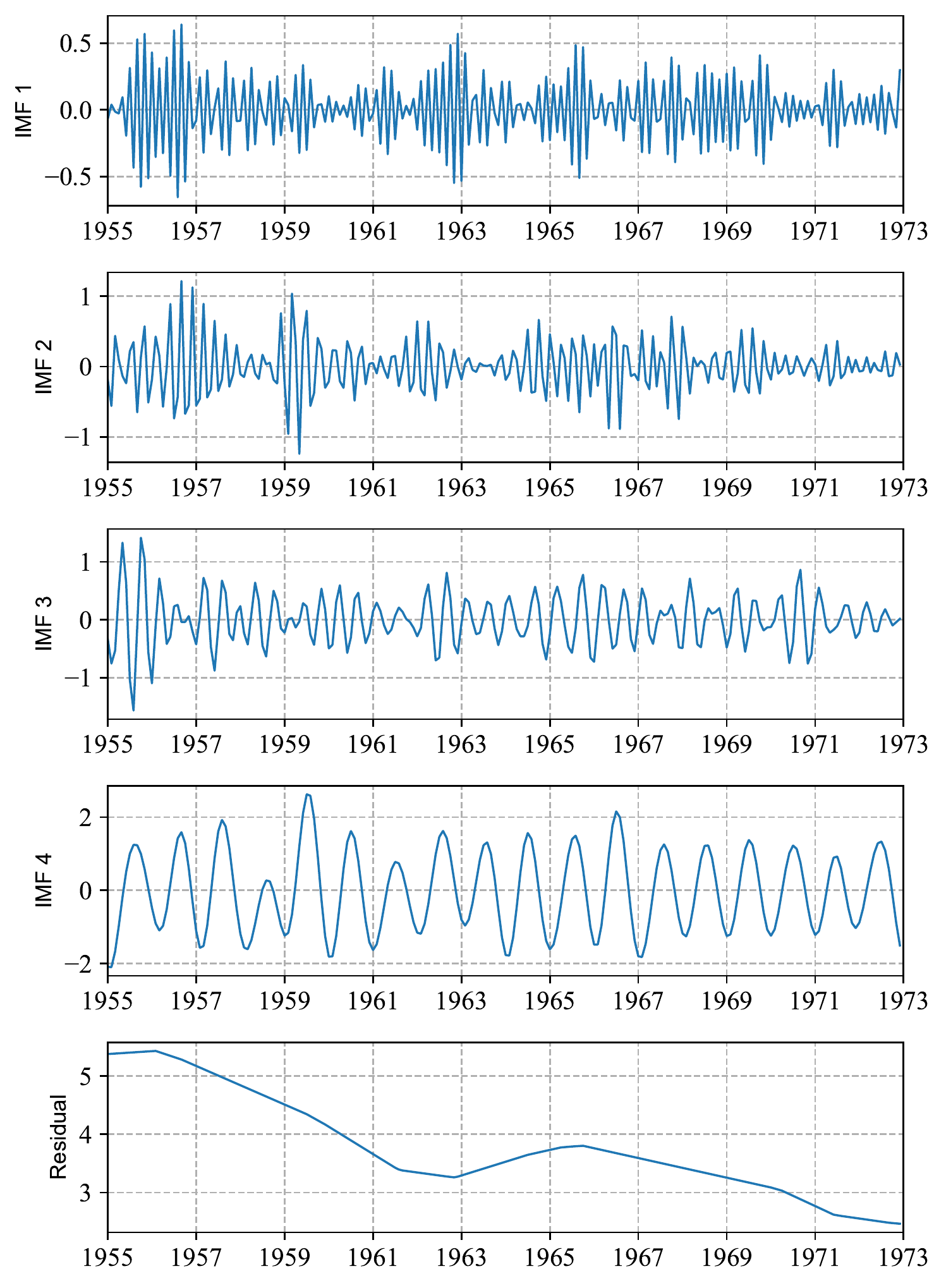}
            \label{Hu15c}
            \vspace{-2em}
			\caption{NMD}
		\end{subfigure}
        \vspace{-1em}
        \caption{Decompositions of EMD, VMD and NMD on Ozone data. (a) EMD's results. (b) VMD's results. (c) NMD's results.}
        \vspace{-1em}
        \label{Hu15}
	\end{figure}

\emph{4) Ozone Concentration data}: This dataset is the monthly ozone concentration in downtown Los Angeles from January 1955 to August 1967 \cite{hipel1994time}, as illustrated in Fig. \ref{Hu14}, and the decomposition results of EMD, VMD and NMD are shown in Fig. \ref{Hu15}.

Ozone data has a strong noise and more random components than Airline data, which bring some difficulties on components recovering. EMD do not recover any complete frequency components except the long-term trend and do not present any separate mode of high frequency in Fig. \ref{Hu15}. The frequency of each mode of VMD and NMD is the same, but differ a lot on amplitude. There is significant noise exists in high frequency IMF of NMD. In contrary, we do not see any obvious noise in VMD results, suggesting that VMD do a better job on noise filtering. We list the PE and correlation coefficient of modes and original data on Ozone data in Table. \ref{T2}. The VMD and NMD's correlation coefficient results are really similar, but their PE have significant differences, in which PE of VMD's IMFs are all more than NMD's, showing that the noise exists in different parts of VMD and NMD results. This phenomenon may suggest that the noise would not effects the correlation coefficient. The most meaningful periods of Ozone concentration are 6 and 12 months, reflected in IMF3 and IMF4 of VMD and NMD.

\begin{table}
\caption{PE and Corr of modes and original data on Ozone dataset.}
\vspace{-1em}
\label{T2}
\begin{minipage}{\columnwidth}
\begin{center}
\begin{tabular}{ccccccc}
\hline
\multicolumn{2}{c}{}& IMF1&IMF2&IMF3&IMF4&Res\\\hline
 \multirow{2}*{EMD} &PE&2.73&7.64&1.42&1.09&87.10\\
\cline{2-7}
~&Corr&0.40&0.57&0.12&0.24&0.51\\\hline
 \multirow{2}*{VMD}&PE&0.29&0.51&0.67&6.56&91.96\\
\cline{2-7}
~&Corr&0.20&0.27&0.30&0.72&0.58\\\hline
 \multirow{2}*{NMD}&PE&0.40&0.84&1.07&7.06&90.62\\
\cline{2-7}
~&Corr&0.17&0.25&0.29&0.72&0.54\\\hline
\end{tabular}
\end{center}
\vspace{-1em}
\end{minipage}
\end{table}

 \begin{table*}
\caption{IO and MAE of EMD, VMD and NMD on 4 datasets.}
\vspace{-1em}
\label{T3}
\begin{center}
\begin{tabular}{ccccccccc}
\hline
\multirow{2}*{}& \multicolumn{2}{c}{$x_{1}(t)$}&\multicolumn{2}{c}{$x_{2}(t)$}&\multicolumn{2}{c}{Airline}&\multicolumn{2}{c}{Ozone}\\
\cline{2-9}
~&IO&MAE&IO&MAE&IO&MAE&IO&MAE\\\hline
 EMD&6.1$\times10^{-3}$&5.7$\times10^{-17}$&-1.4$\times10^{-4}$&1.9$\times10^{-17}$&-2.5$\times10^{-3}$&3.4$\times10^{-9}$&-4.7$\times10^{-2}$&2.4$\times10^{-16}$\\\hline
 VMD&7.7$\times10^{-3}$&1.2$\times10^{-2}$&3.7$\times10^{-3}$&4.1$\times10^{-2}$&1.2$\times10^{-4}$&1.7$\times10^{5}$&4.1$\times10^{-3}$&8.8$\times10^{-2}$\\\hline
 NMD&6.7$\times10^{-4}$&3.4$\times10^{-3}$&-5.7$\times10^{-2}$&1.5$\times10^{-3}$&-1.3$\times10^{-3}$&9.3$\times10^{3}$&-4.2$\times10^{-3}$&4.2$\times10^{-3}$\\\hline
\end{tabular}
\end{center}
\vspace{-1em}
\end{table*}

\begin{figure}
		\begin{subfigure}[t]{0.2\textwidth}
			\centering
			\includegraphics[width=\textwidth]{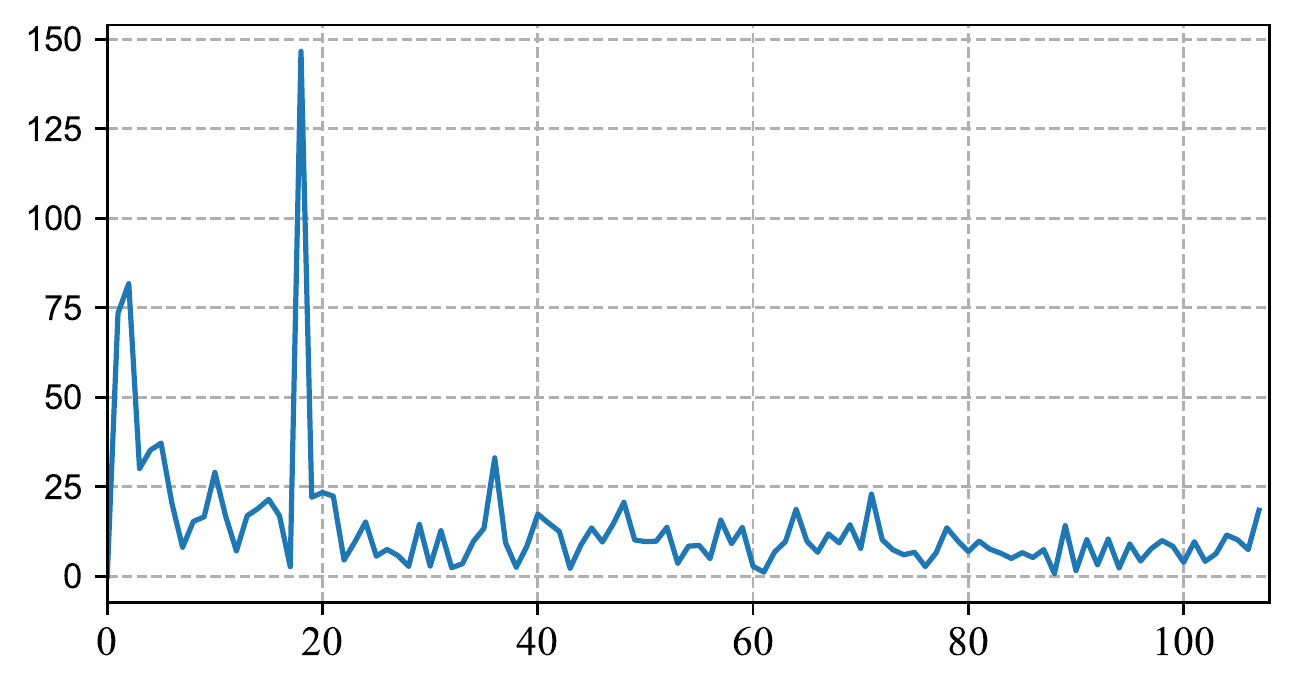}
            \label{Hu16a}
            \vspace{-2em}
			\caption{Ozone data}
		\end{subfigure}
		\begin{subfigure}[t]{.2\textwidth}
			\centering
			\includegraphics[width=\textwidth]{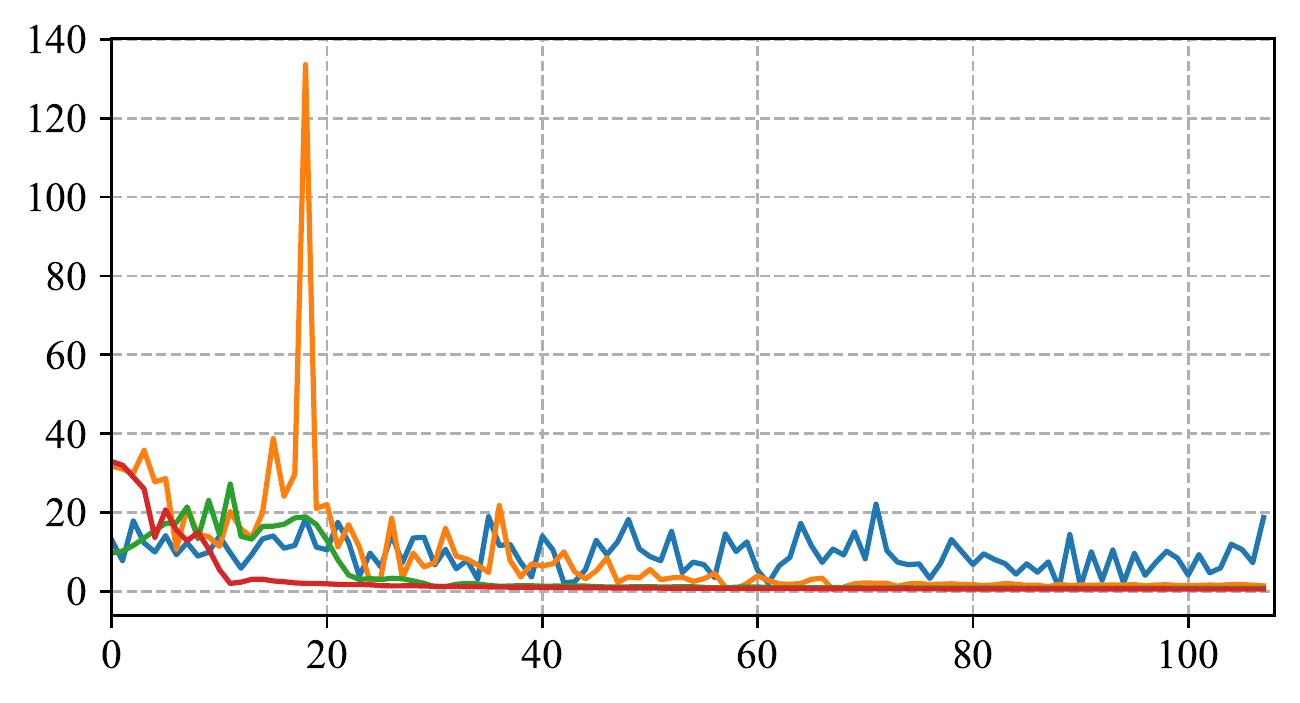}
            \label{Hu16b}
            \vspace{-2em}
			\caption{EMD}
		\end{subfigure}
		\begin{subfigure}[t]{.2\textwidth}
			\centering
			\includegraphics[width=\textwidth]{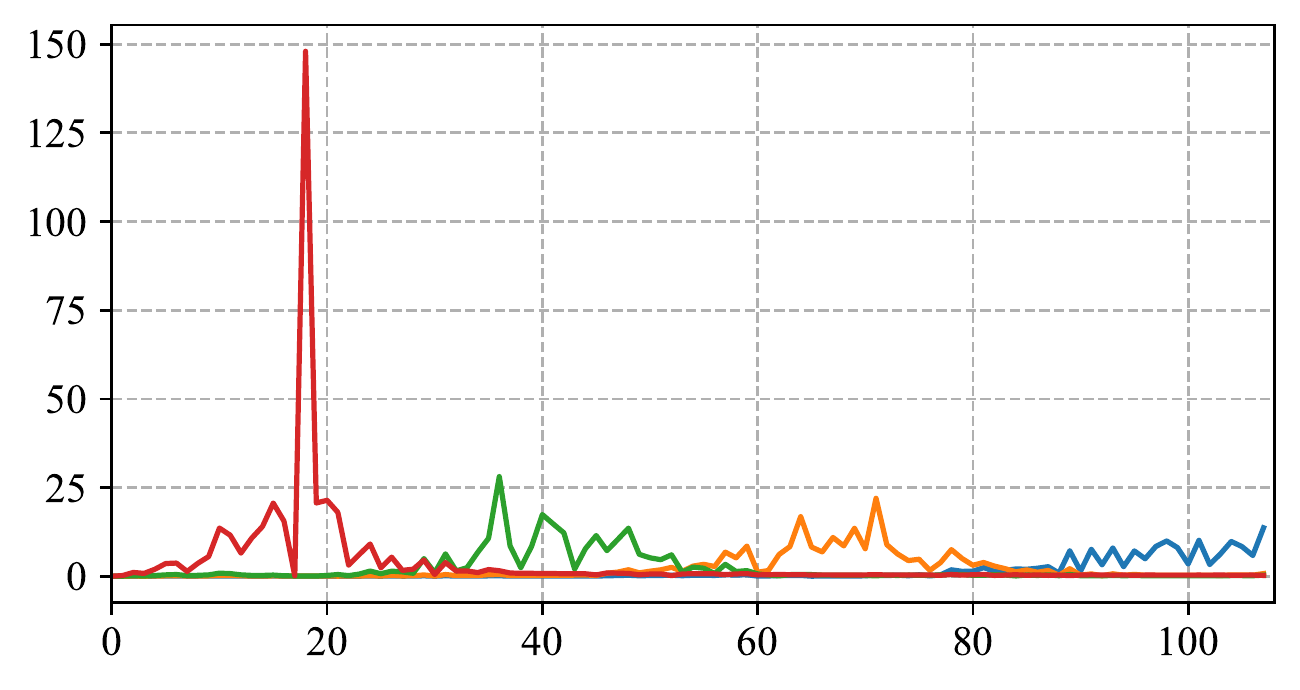}
            \label{Hu16c}
            \vspace{-2em}
			\caption{VMD}
		\end{subfigure}
		\begin{subfigure}[t]{.2\textwidth}
			\centering
			\includegraphics[width=\textwidth]{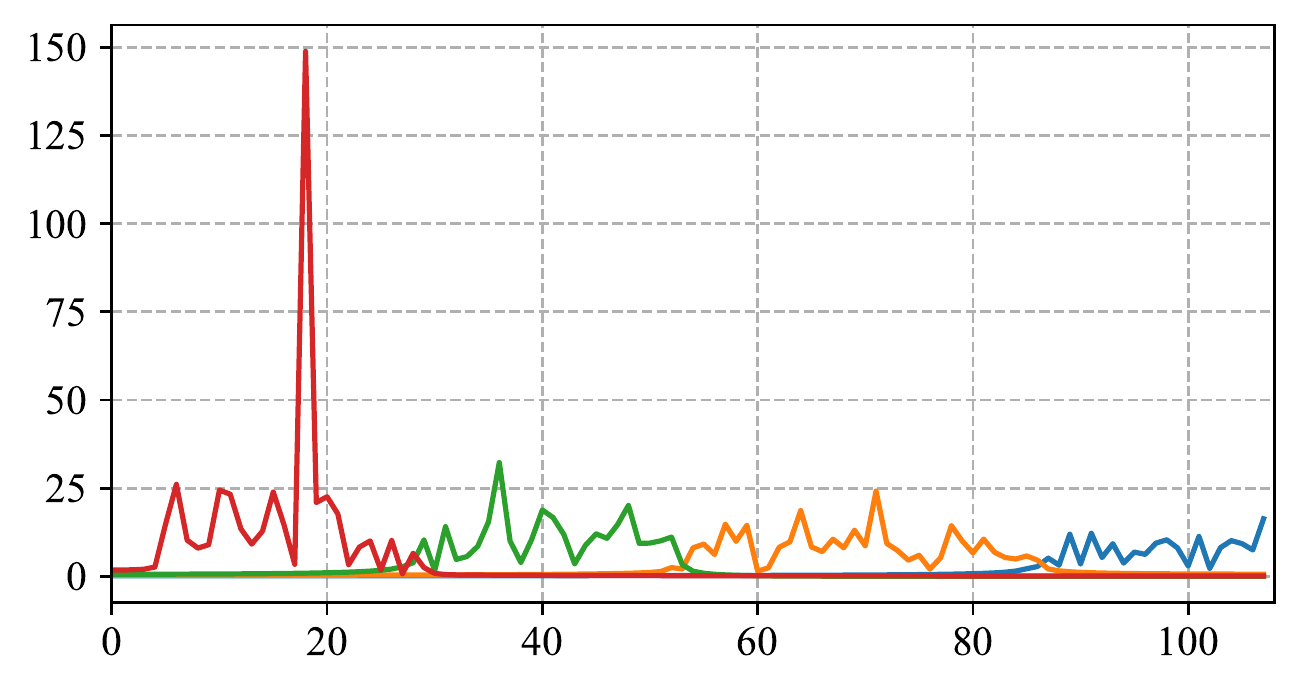}
            \label{Hu16d}
            \vspace{-2em}
			\caption{NMD}
		\end{subfigure}
        \vspace{-1em}
        \caption{Spectrum of Ozone data and its decompositions. (a) Spectrum of zero-mean normalized Ozone data. (b) Spectrum of EMD results. (c) Spectrum of VMD results. (d) Spectrum of NMD results.}
        \vspace{-1em}
        \label{Hu16}
	\end{figure}

Fig. \ref{Hu16} illustrate the DFT of EMD, VMD and NMD results. In Fig. \ref{Hu16}, EMD results still have serious frequency mixing between each mode on this data. The DFT of VMD modes are compactly supported at the central frequency, and the frequency components far from the central frequency do not appear in IMF spectrum but in long-term trend. This is the results of noise filtering, but may cause VMD not reflecting some relatively small but important frequency components. The NMD results can completely separate the frequency domain without overlapping, and the random components of data will be evenly separated in each IMF. NMD can fully reflect all the frequency components of the original data, but its ability of filtering noise is weak.


\subsubsection{Analysis of orthogonality and completeness}\label{3.2.2}
{}

Orthogonality has many advantages on signal processing from the meaning of mathematics, so we often need the decomposed components are orthogonal, while EMD, VMD and NMD have only approximate orthogonality without strict proof of theoretical orthogonality. Hence, we test its orthogonality by experiments with the overall index of orthogonality (IO) defined in \cite{huang1998empirical}. The closer IO is to 0, the better orthogonality the decomposition results have. We rewrite the formula of IO as follows,
\begin{equation}\label{eq5}
IO=\frac{\sum^{N}_{t=1}\sum^{K+1}_{i,j,i\neq j}u_{i}(t)u_{j}(t)}{\sum_{t=1}^{N}x^2(t)}.
\end{equation}

There may be some errors between original data and reconstructed data. In EMD, the reconstruction error is caused by the calculation accuracy; In VMD, the calculation accuracy is caused by the degree of convergence; In NMD, the calculation accuracy is caused by the training of neural network. We use mean absolute error (MAE) between the sum of decomposed modes and original data to measure reconstruction error of three decomposition methods

We calculate IO and MAE of EMD, VMD and NMD on 4 datas introduced in Section. \ref{sec3.2.1}, whose results are listed in Table. \ref{T3}. The IO of NMD is the best on $x_1(t)$ and the worst on $x_2(t)$, and has a similar performance with VMD on airline passengers and ozone concentration datas, better than EMD. EMD shows a best orthogonality on $x_2(t)$ and a worst on ozone concentration data, and VMD shows a best orthogonality on $x_2(t)$ and airline passengers data and a worst on $x_1(t)$. A generous review of this results is that the orthogonality of these three algorithm are all a bit temperamental, and VMD perform relatively the best, followed by NMD and EMD in turn. The MAE of original data and the sum of decomposed components of EMD is always the least and can be viewed as 0, indicating that EMD has full completeness. NMD has a relatively larger but still small enough MAE, also showing a good completeness. VMD has the worst completeness, which may cause some misleading interpretation on further analysis of data.

\subsection{Extrapolation}\label{sec3.3}
\begin{figure}[htbp]
		\begin{subfigure}[t]{0.23\textwidth}
			\centering
			\centerline{\includegraphics[width=\textwidth]{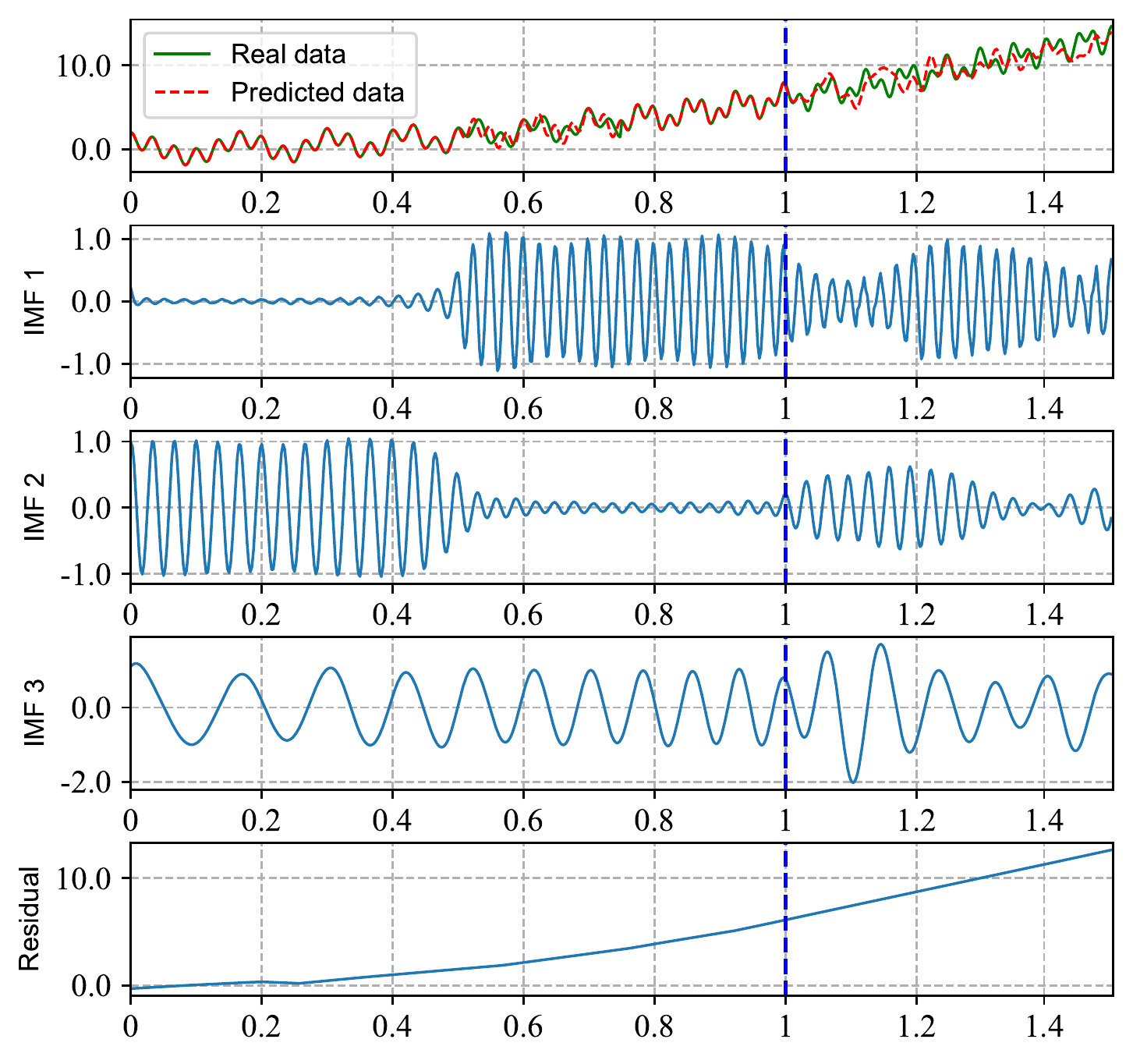}}
            \label{Hu17a}
            \vspace{-0.7em}
			\caption{$x_{1}(t)$}
		\end{subfigure}
        \hfill
		\begin{subfigure}[t]{.23\textwidth}
			\centering
			\centerline{\includegraphics[width=\textwidth, height=1.2in]{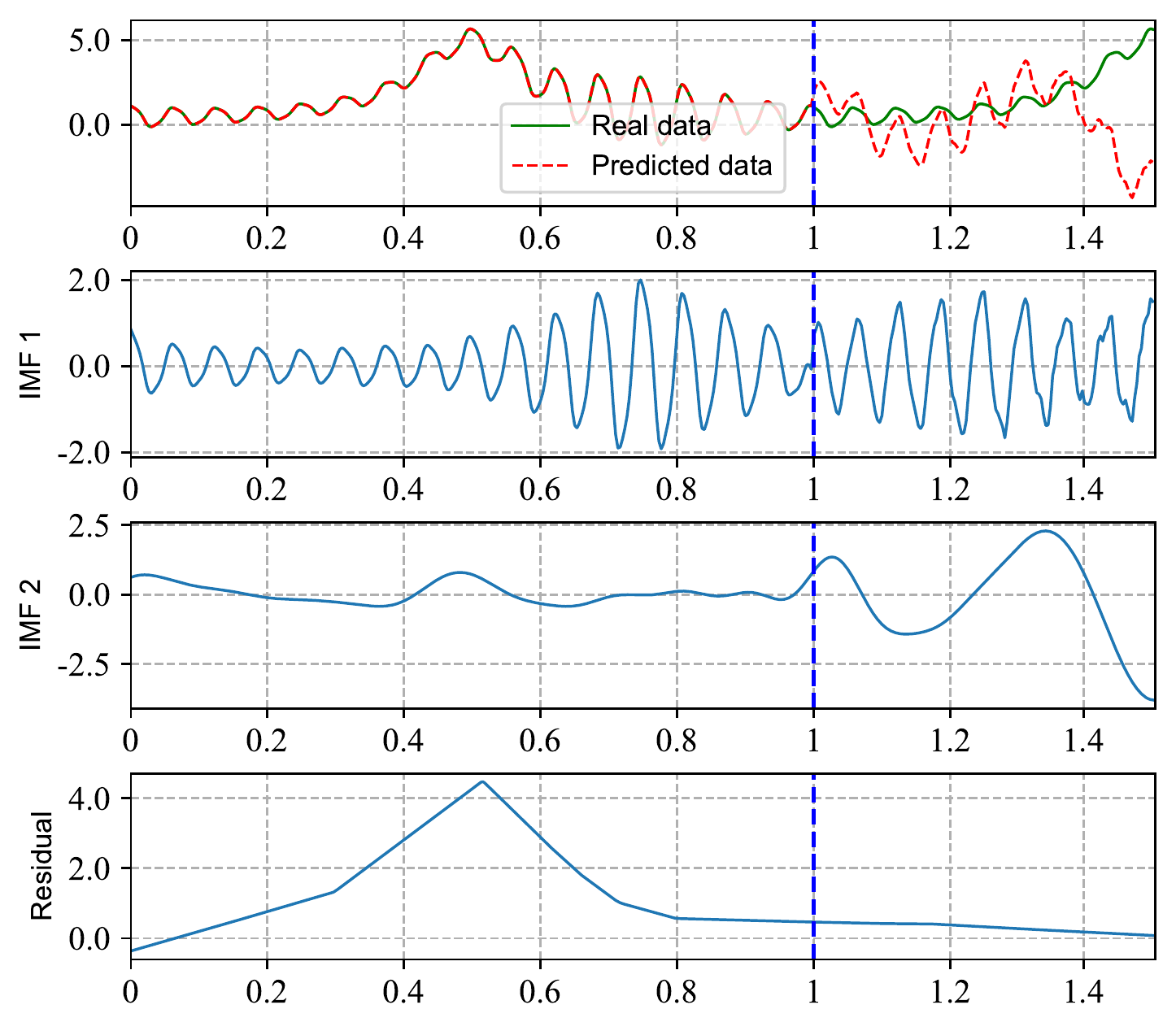}}
            \label{Hu17b}
            \vspace{-0.7em}
			\caption{$x_{2}(t)$}
		\end{subfigure}
        \hfill
		\begin{subfigure}[t]{.235\textwidth}
			\centering
			\centerline{\includegraphics[width=\textwidth]{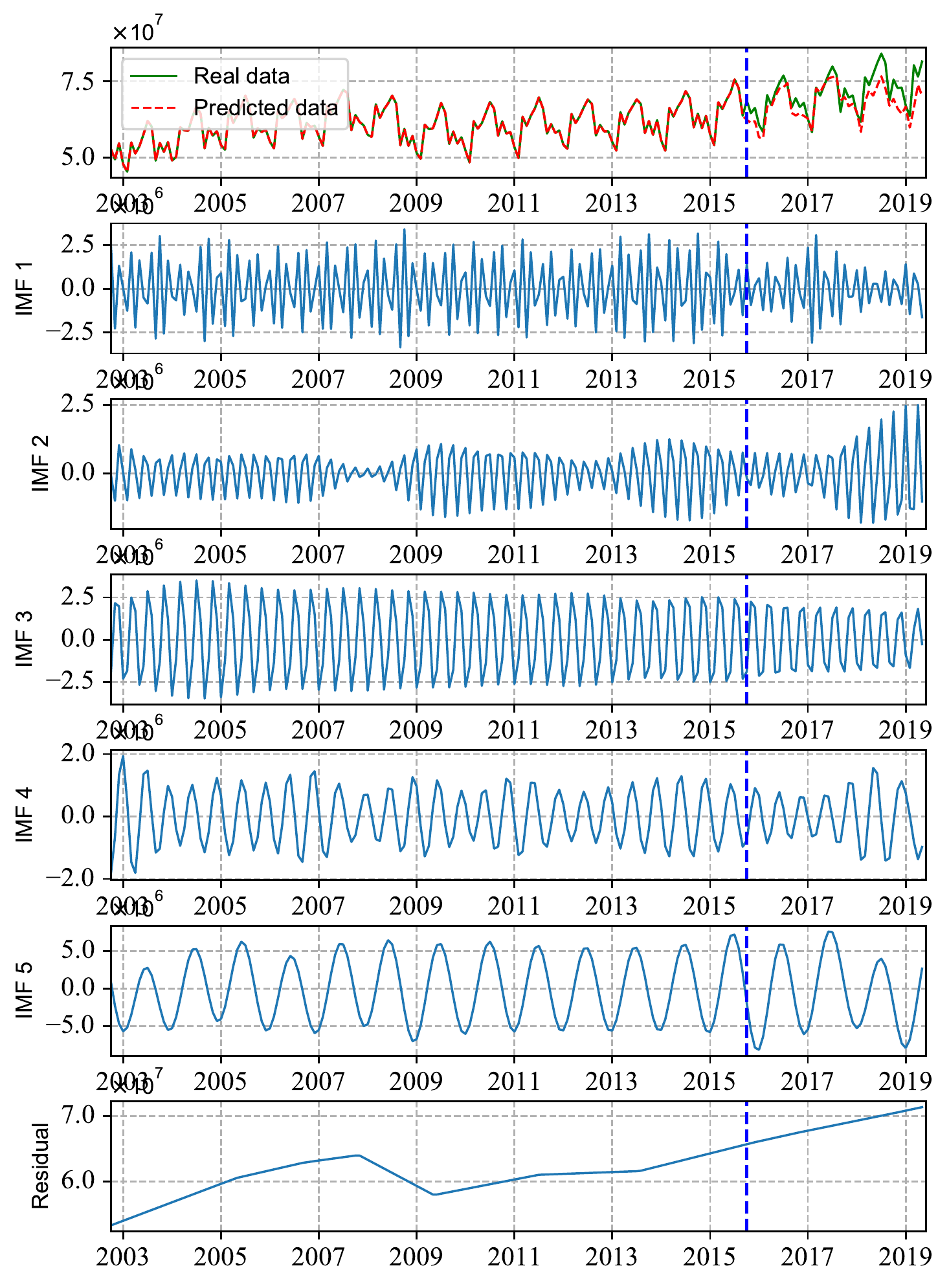}}
            \label{Hu17c}
            \vspace{-0.7em}
			\caption{Airline data}
		\end{subfigure}
        \hfill
		\begin{subfigure}[t]{.235\textwidth}
			\centering
			\centerline{\includegraphics[width=\textwidth]{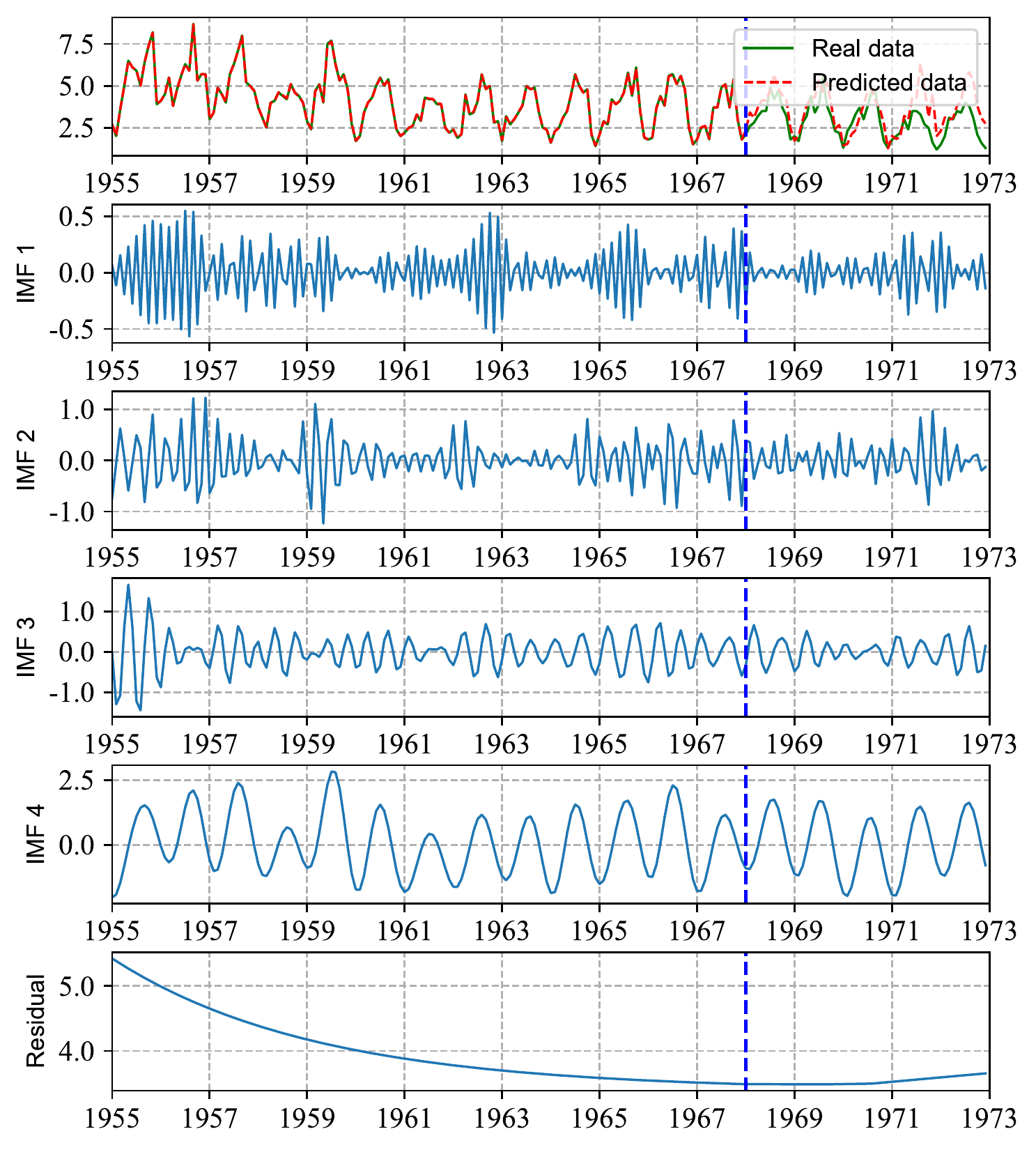}}
            \label{Hu17d}
            \vspace{-0.7em}
			\caption{Ozone data}
		\end{subfigure}
        \vspace{-1em}
        \caption{Extrapolation of FNN on 4 datas. (a) Extrapolation of $x_{1}(t)$ and its components. (b) Extrapolation of $x_{2}(t)$ and its components. (a) Extrapolation of Airline data and its components. (a) Extrapolation of Ozone data and its components.}
        \vspace{-1.5em}
        \label{Hu17}
	\end{figure}

In the process of training FNN, we can obtain a exact equation between data and time from, which makes the FNN have the potential capability of extrapolation and prediction. The extrapolation of the fitted function can reflect the authenticity and effectiveness of variable relations, which can give us a different point of view on feature mining. In this section, the variables on the extrapolation interval are inputed into the trained FNN to study its extrapolation on 4 datas introduced in Section \ref{sec3.2}. The extrapolation results on complete data and decomposed components of NMD are illustrated in Fig. \ref{Hu17}.

In Fig. \ref{Hu17}, the green curve represents the real data, and the red curve represents the output of FNN. The curve before the blue dotted line is on training interval, and the curve after the blue dotted line is on extrapolation interval. From the extrapolation of decomposed components of $x_{1}(t)$, we can infer that the extrapolation errors result from the end effect, which bring some trend deviation on frequency amplitude at the end of training interval. The results on $x_{2}(t)$ show that, due to the training interval only contains one period of original data, this insufficient sampling leads to FNN collecting a wrong frequency feature from training. In Airline data, the prediction errors comes from the trend. The sub-network of nonperiodic component in FNN utilize Relu, which cause the trend growing with linear in extrapolation interval, differing from the actual nonlinear trend. The causes of prediction errors in Ozone data are the same as Airline data.

Overall, it can be concluded that when the fitted function extrapolate the data, the results are very sensitive to the trend at the end of training interval, and the trend deviation increase with the length of extrapolation interval. The sufficient length of training interval is also important for neural network to learn the real periodic characteristics. Predicting data only utilizing the relation between time domain and sample data has a weak robustness and insufficient information. It needs to be combined with other features such as last several sampling datas to achieve a better prediction.

\section{Conclusion}\label{sec4}
This paper propose a mode decomposition method NMD based on FNN that provides a decomposition algorithm with good adaptability and mathematical theory. The algorithm affords to decompose the data into a group of AM signals utilizing FNN, and fit the relation between time and data at the same time, so that we can analyze the decomposition results from the perspective of extrapolation results. The developed clustering algorithm can extract the main frequency components from the AM signals, and divide all AM signals into various modes centered on the main frequency components. The experiments show NMD can efficiently implement mode decomposition. The components decomposed by NMD have a good performance on feature collecting and end effect robustness, has a better noise immunity than EMD and has a better completeness than VMD.

The strong adaptability of NMD comes from the technique of neural network, but which may also bring some instability on decomposition results, and consumes more times. In the future work, we will explore the application of other neural networks in mode decomposition, and further investigate the intrinsic features of frequency in datas.
\bibliographystyle{ACM-Reference-Format}
\bibliography{NMD}

\end{document}